\documentclass[fleqn,usenatbib]{mnras}

\usepackage{newtxtext,newtxmath}
\usepackage{mathrsfs}

\usepackage[T1]{fontenc}

\DeclareRobustCommand{\VAN}[3]{#2}
\let\VANthebibliography\thebibliography
\def\thebibliography{\DeclareRobustCommand{\VAN}[3]{##3}\VANthebibliography}

\usepackage{graphicx}	% Including figure files
\usepackage{amsmath}	% Advanced maths commands
\usepackage{amssymb}	% Extra maths symbols

\newcommand{\beq}{\begin{equation}}
\newcommand{\eeq}{\end{equation}}
\newcommand{\beqa}{\begin{eqnarray}}
\newcommand{\eeqa}{\end{eqnarray}}

\newcommand{\bc}{\begin{center}}
\newcommand{\ec}{\end{center}}

\newcommand{\peryr}          {\,{\rm yr}^{-1}}
\newcommand{\Myr}            {\,{\rm Myr}}

\newcommand{\Gyr}            {\,{\rm Gyr}}

\newcommand{\kpc}            {\,{\rm kpc}}
\newcommand{\Mpc}            {\,{\rm Mpc}}
\newcommand{\Msun}           {\,{\rm M}_\odot}
\newcommand{\Msunyr}         {\,{\rm M}_\odot\,{\rm yr}^{-1}}

\newcommand{\hMpc}           {\,h^{-1}\,{\rm Mpc}}

\newcommand{\kms}            {\,\,{\rm km}\,\,{\rm s}^{-1}}
\newcommand{\kmsMpc}         {\,\,{\rm km}\,\,{\rm s}^{-1}\,{\rm Mpc^{-1}}}

\newcommand{\YunnanII}       {{\slshape Yunnan-II}}

\newcommand{\Meject}          {M_{\rm eject}}

\newcommand{\Om}              {\Omega_{\rm m}}
\newcommand{\Ob}              {\Omega_{\rm b}}
\newcommand{\OL}              {\Omega_{\rm \Lambda}}
\newcommand{\seight}          {\sigma_{\rm 8}}
\newcommand{\Hzero}           {H_{\rm 0}}

\newcommand{\CODE}             {{GABE}}

\newcommand{\RCOM}           {R_{\rm COM}} % \mathbf{R} == !19R!X in idl
\newcommand{\Rkilonova}       {R_{\rm kilonova}}
\newcommand{\RNNM}            {R_{\rm NNM}}
\newcommand{\RNBM}            {R_{\rm NBM}}
\newcommand{\RsspCOM}         {\mathcal{R}_{\rm ssp,COM}}
\newcommand{\RsspNNM}         {\mathcal{R}_{\rm ssp,NNM}}
\newcommand{\RsspNBM}         {\mathcal{R}_{\rm ssp,NBM}}
\newcommand{\tage}            {t_{\rm age}}
\newcommand{\Vkick}           {V_{\rm kick}}
\newcommand{\nCOM}            {n_{\rm COM}}
\newcommand{\nNNM}            {n_{\rm NNM}}
\newcommand{\nNBM}            {n_{\rm NBM}}
\newcommand{\fOld}            {f_{\rm Old}}
\newcommand{\PCOM}            {P_{\rm COM}}
\newcommand{\PNNM}            {P_{\rm NNM}}
\newcommand{\PNBM}            {P_{\rm NBM}}

\newcommand{\fBT}             {f_{\rm B/T}}

\newcommand{\Mrp}             {M_{\rm rp}}
\newcommand{\NlifeNNM}        {N_{\rm life,NNM}}
\newcommand{\NlifeNBM}        {N_{\rm life,NBM}}
\newcommand{\MejectaNNM}     {M_{\rm ejecta,NNM}}
\newcommand{\MejectaNBM}     {M_{\rm ejecta,NBM}}
\newcommand{\MrpAs}          {M_{\rm rp,A>79}}

\newcommand{\Mstar}           {M_{\rm *}}

\newcommand{\dif}             {\mathrm{d}}
\newcommand{\Gpcyr}           {\,\,{\rm Gpc}^{-3}\,\,{\rm yr}^{-1}}

\title[Simulating Kilonovae in the $\Lambda$CDM Universe]{Simulating Kilonovae in the $\Lambda$CDM Universe}

\author[Z. Jiang et al.]{
\parbox{\textwidth}{Zhen Jiang,$^{1,2}$\thanks{E-mail: zjiang@nao.cas.cn}
Jie Wang,$^{1,2}$
Fenghui Zhang,$^{3,4}$
Li-Xin Li,$^{5}$
Lan Wang,$^{1}$
Ran Li,$^{1,2}$\\
Liang Gao,$^{1,2,7}$
Zhanwen Han,$^{3,4,6}$
and Jun Pan$^{1}$}\vspace{0.4cm}
\\
$^{1}$Key Laboratory for Computational Astrophysics, National Astronomical Observatories, Chinese Academy of Sciences, Beijing 100012, China\\
$^{2}$School of Astronomy and Space Science, University of Chinese Academy of Sciences, Beijing 100049, China\\
$^{3}$Yunnan Observatories, Chinese Academy of Sciences, Kunming, 650216, China\\
$^{4}$Key Laboratory for the Structure and Evolution of Celestial Objects, Chinese Academy of Sciences, Kunming, 650216, China\\
$^{5}$Kavli Institute for Astronomy and Astrophysics, Peking University, Beijing 100871, China\\
$^{6}$Center for Astronomical Mega-Science, Chinese Academy of Sciences, 20A Datun Road, Chaoyang District, Beijing, 100012, China\\
$^{7}$Institute of Computational Cosmology, Department of Physics, University of Durham, South Road,Durham DH1 3LE, UK
}

\date{Accepted XXX. Received YYY; in original form ZZZ}

\pubyear{2020}

\begin{document}
\label{firstpage}
\pagerange{\pageref{firstpage}--\pageref{lastpage}}
\maketitle

\begin{abstract}
Kilonovae are optical flashes produced in the aftermath of neutron star-neutron star mergers (NNMs) or neutron star-black hole mergers (NBMs). The multi-messager observation of the recent gravitational wave event GW170817 confirms that it originated from a NNM and triggered a kilonova. In this work, we use the Millennium Simulation, combined with a semi-analytic galaxy formation model--GABE (Galaxy Assembly with Binary Evolution) which adopts binary stellar population synthesis models, to explore the cosmic event rate of kilonovae, and the properties of their host galaxies in a cosmological context. We find that model with supernova kick velocity of $V_{\rm kick}=0\kms$ fits the observation best, in agreement with the exception of some formation channels of binary neutron star. This indicates that NNMs prefer to originate from binary systems with low kick velocities. With $V_{\rm kick}=0\kms$, the cosmic event rate of NNMs and NBMs at $z=0$ are $283\Gpcyr$ and $91\Gpcyr$, respectively, marginally consistent with the constraint from LVC GWTC-1. For Milky Way-mass galaxies, we predict the NNM rate is $25.7^{+59.6}_{-7.1}\Myr^{-1}$, which is also in good agreement with the observed properties of binary neutron stars in the Milky Way. Taking all the NNMs into account in the history of Milky Way-mass galaxies, we find that the averaged r-process elements yield with $A>79$ in a NNM and NBM event should be $0.01\Msun$ to be consistent with observation. We conclude that NGC 4993, the host galaxy of GW170817, is a typical host galaxy for NNMs. However, generally NNMs and NBMs tend to reside in young, blue, star-forming, late-type galaxies, with stellar mass and gaseous metallicity distribution peaking at $\Mstar=10^{10.65}\Msun$ and $12+\log{\rm (O/H)}=8.72-8.85$, respectively. By studying kilonovae host galaxies in the cosmological background, it is promising to constrain model details better when we have more events in the forthcoming future.
\end{abstract}

\begin{keywords}
galaxies: general -- binaries: close -- neutron star mergers -- black hole - neutron star mergers -- supernovae: general -- nuclear reactions, nucleosynthesis, abundances
\end{keywords}

%%%%%%%%%%%%%%%%%%%%%%%%%%%%%%%%%%%%%%%%%%%%%%%%%%

%%%%%%%%%%%%%%%%% BODY OF PAPER %%%%%%%%%%%%%%%%%%
\section{Introduction}
\label{section:intro}
When a neutron star-neutron star merger (NNM) happens, neutron-rich material is ejected subrelativistically and a black hole or a neutron star is left over as a remnant \citep{Abbottremnant,Yu18}. The neutron-rich expanding ejecta provides an excellent nursery for rapid neutron capture (r-process) nucleosynthesis. The decay radiation of these newly-formed r-process elements is the so-called kilonova \citep{Li98,Metzger10,Metzger17}, which is expected to appear days after merger and peak at ultraviolet, optical, or near-infrared wavelengths, depending on the opacity of ejecta \citep{Li98,Kasen13,Tanaka13,Barnes13}. Kilonova was first directly observed through infrared emission excess about one week after SGRB 130603B \citep{Tanvir13,Berger13}. Note that theoretically, neutron star-black hole mergers (NBMs) can also eject neutron-rich matter with sub-relativistic velocity and trigger kilonovae \citep[e.g.][]{Lattimer74,Surman08}. 

On August 17th, 2017, the first directly detected NNM, GW170817, was observed by advanced LIGO detectors \citep{LIGO15,Abbott17}. This was the only NNM event observed in the first (O1) and second (O2) observing run of advanced LIGO (O1 spanned four months, and O2 spanned nine months).  Other detection were all black hole-black hole mergers (BBMs), with 10 confidently identified detection \citep{LIGO18}. No NBMs were detected. Considering all available data from O1 and O2, LIGO Scientific Collaboration and Virgo Collaboration (LVC) infer that the cosmic event rate of NNMs is $1210_{-1040}^{+3230}\Gpcyr$ with $90\%$ confidence, and the $90\%$ upper limit of the cosmic event rate of NBMs is $610\Gpcyr$ \citep{GWTC1}.

%The estimated cosmic event rate of NNMs is $1540^{+3200}_{-1220}\Gpcyr$, with the huge error caused by the limited number of NNM detected.

Not only was GW170817 detected in gravitational waves (GWs), its counterparts in $\gamma$-ray, X-ray, UV, optical, infrared, and radio bands were also recognized in the sky area constrained by advanced LIGO and advanced Virgo \citep[summarized in ][]{Abbott17multi}. Its electromagnetic (EM) emission peaked $<1{\,\rm day}$ in ultraviolet, indicating a blue component with low opacity, and then slowly shifted towards near-infrared days after merger, which can be fitted with a two-component kilonova model \citep{Tanvir17,Cowperthwaite17,Nicholl17,Waxman18,Li19}. With much higher accuracy on localization than GW detection, these multi-bands observations confirmed that NGC 4993 is the host galaxy of GW170817. NGC 4993 is an old elliptical galaxy with stellar mass of $\log{(\Mstar/\Msun)}=10.65^{+0.03}_{-0.03}$ and median mass-weighted age of $13.2^{+0.5}_{-0.9}\Gyr$ as measured by \cite{Blanchard17}. \cite{Troja17} reported a similar stellar mass, $\log{(\Mstar/\Msun)}=10.7-11.0$, but with a younger age of $3-7\Gyr$. As the spectra of long-lived stars evolve quite slowly, the age estimation of stellar populations of old galaxies has large systematic errors. 

The third (O3) observing run of LVC\footnote{https://gracedb.ligo.org/superevents/public/O3/} began on 1st April, 2019 and is planned to end on 30th April, 2020. Till the end of 2019, with nine months' observation, three NNMs and two NBMs (probability $>99\%$) have been detected. Among them, S190425z \citep{S190425z} has a probability $>99\%$ to be a NNM, with a false alarm rate (FAR) of 1 per 69834 years. $\sim0.5$s and $\sim5.9$s later, a weak $\gamma$-ray burst which consisted of two pulses was detected by INTEGRAL, in the northern region of the localization proposed by LVC \citep{Pozanenko19}. However, following observations have not yet confirmed any optical counterpart of S190425z \citep{MMT190425z,SAGUARO190425z,GRWOTH190425z,GRANDMA190425z}. S190814bv\footnote{https://gracedb.ligo.org/superevents/S190814bv/view/} has a probability $>99\%$ to be a NBM, with an extremely low FAR of 1 per $1.559\times10^{25}$ years. Unfortunately, no EM counterparts were confirmed in the following observing campaign till the end of 2019 \citep{Gomez19,Andreoni19,Dobie19}. Information about candidates and host galaxies of other events are not public yet. As O3 is still going on and its updates on NNMs and NBMs event rate and host galaxies have not been published, we stick to the observational result of LVC O1 and O2 as the comparison with our model prediction in this work.

%However, so far there is no direct observation of NBM events, neither are their associated kilonovae.

%GW170817 has brought the community two observational facts: the cosmic event rate of NNMs is $1540^{+3200}_{-1220}\Gpcyr$; and the first NNM event is observed in an old elliptical galaxy. 
%It is not clear yet whether these findings related to GW170817 are consistent with the predictions of current stellar evolution and galaxy formation theories.
From the modelling point of view, the traditional way to estimate the NNM event rate in stars of a certain galaxy is to convolve the NNM event rates of simple stellar populations which are derived from stellar population synthesis models with a hypothetical star formation history. For instance, based on observational results \citep[e.g.][]{Gilmore01}, a constant star formation rate of $3.5-4.0\Msunyr$ that lasts for $10-12\Gyr$ is usually assumed to estimate the NNM event rate in the Milky Way \citep{Portegies98,Belczynski02,Voss03,Belczynski07,Dominik12,Belczynski16,Chruslinska18,Belczynski18,Belchannels18}. The cosmic event rate density can then be derived by considering the number density of
Milky Way-type galaxies \citep{Belczynski07,Belczynski16}. The cosmic event rate density can also be estimated \citep{Dominik13,Chruslinska18,Boco19} using fitting formulas of the history of cosmic star formation density \citep{Strolger04,Madau14}. In recent years, due to the fast development of cosmological hydrodynamic simulations, star formation histories from cosmological hydrodynamic simulations are more frequently used to estimate NNM event rate \citep{Mapelli18,Mapelli182,Mapelli19,Toffano19,Artale192,Artale19}. Star formation histories of various galaxies derived from semi-empirical models \citep{Behroozi18} are also used \citep{Adhikari20}.
%Though all these star formation histories are derived based on observations, they are not self-consistent in between galactic scale and cosmic scale. Besides, it is tedious to analyze the properties of NNMs' host galaxy in this case study-like method. Whether NGC 4993 is a typical host galaxy can only be answered after comparing all kinds of galaxies, i.e. all kinds of star formation histories.

In this work, we use semi-analytic models of galaxy formation to estimate the cosmic event rate of kilonova events, triggered by both NNMs and NBMs (hereafter denoted as compact object mergers, COMs), and study the properties of their host galaxies. Combined with N-body merger trees of dark matter haloes, semi-analytic models trace how galaxies form and evolve in haloes, by implementing simplified models or empirical relations that describe physical processes including reionization, gas cooling, star formation, supernova feedback, black hole growth, AGN feedback, galaxy mergers etc, and have recovered a large amount of observations in the local universe and at high redshift  \citep[e.g.][]{White91,Kauffmann99,Croton06,DeLucia07,Guo11,Henriques15}. With semi-analytic models, the star formation histories of galaxies in a large mass range, from dwarf satellite galaxies to BCGs (the Brightest Cluster Galaxy), are specified from first principle, which allows us to derive NNM and NBM event rates for each galaxy, and to explore the relationship between kilonovae and their host galaxies. Besides, semi-analytic models consume much less computational time than cosmological hydrodynamic simulations, which allows us to generate galaxy catalogue for a larger volume.
%could be helpful for further detection of NBMs and their kilonovae.
%use prescriptions based method, which are either derived from simplified galactic models or empirical relations from observations, to build a rather complete cycle for different components in the evolution of galaxies. From early 1990s to 2010s, semi-analytic models have incorporated many galactic processes, including reionization, gas cooling, star formation, supernova feedback, black hole growth, AGN feedback, disk instability, metal enrichment, dynamical friction, satellite stripping, galaxy mergers, etc, and have recovered a large amount of observations in the local universe and even at high redshift (e.g. \cite{White91}; \cite{Kauffmann99}; \cite{Croton06}; \cite{DeLucia07}; \cite{Guo11}; \cite{Henriques15}). Combined with N-body merger trees, semi-analytic models can simultaneously recover properties of galaxies in a large mass range, from dwarf satellite galaxies to BCGs (the Brightest Cluster Galaxy). All these galaxies' evolution are self-consistent and provide us a star formation history library with extremely large diversities. By replacing the hypothetical or statistical star formation histories mentioned before with this library, we can simultaneously derive NNM event rates for all kinds of galaxies, which allows us to explore the relationship between NNM events and their host galaxies.

The semi-analytic model we use in this study is {\CODE} \citep[Galaxy Assembly with Binary Evolution;][]{Jiang19}, which includes a full set of galaxy formation recipes and has reproduced a large body of observational results. Compared with previous semi-analytic models, {\CODE} for the first time modelled binary star evolution by adopting {\YunnanII} stellar population synthesis model, which includes various interactions of binaries.
Therefore, it is able to use {\CODE} to make direct predictions of binary population in the simulated galaxies. In particular, the remnants of binary stars, i.e. all kinds of double compact objects, including double neutron stars (NS-NS), neutron star-black hole (NS-BH) and double black holes (BH-BH), can be modelled and predicted in detail.
%The technical details of {\CODE} and Yunnan-II model will be described in section \ref{section:semi} and section \ref{section:SPSM} respectively.

The structure of this paper is as follows. In section \ref{section:method}, we first introduce briefly the semi-analytic model {\CODE} and {\YunnanII} stellar population synthesis model we use, then describe our method to calculate NNM and NBM event rates in simple stellar population and in galaxies. In section \ref{section:Rssp} and \ref{section:cosmic}, we show the event rates of NNMs and NBMs for both simple stellar population and for modeled galaxies in a cosmological point of view. Section \ref{section:rprocess} shows the prediction of r-process elements produced by COMs in the lifetimes of galaxies. In section \ref{section:host},  properties of COM host galaxies are presented. We summarize our conclusions in section \ref{section:conclusion}.

\section{Models and methods}
\label{section:method}

\subsection{Semi-analytic model}
\label{section:semi}

The semi-analytic galaxy formation model used in this work is  {\CODE} \citep[Galaxy Assembly with Binary Evolution,][]{Jiang19},
%, which considers a full set of known physical models for galaxy formation, including reionisation, gas cooling, star formation, supernova feedback, black hole growth, AGN feedback, stripping of satellites and galaxy mergers. 
which includes detailed modelling of binary star evolution by adopting  {\YunnanII} stellar population synthesis model (introduced later in section \ref{section:SPSM}).  More details about the model can be found in \cite{Jiang19}. %Here we only describe the details related to the calculation of $\RCOM$.

The Millennium Simulation \citep{Springel05} is used to implement {\CODE} in this work. The cosmological parameters adopted are: $\Om$(matter density) $ = 0.25$, $\Ob$(baryon density) $ = 0.045$, $\OL$(dark energy density) $ = 0.75$, $n$(spectral index) $ = 1$, $\seight$(linear predictions for the amplitude of fluctuations within 8$\hMpc$) $ = 0.9$ and $\Hzero$(Hubble constant) $ = 73 \kmsMpc$, derived from a combined analysis of the 2dFGRS \citep{Colless01} and the first-year WMAP data \citep{Spergel03}. Dark matter haloes and subhaloes in the simulation are identified with a friends-of-friends group finder \citep{Davis85} and SUBFIND \citep{Springel01}, respectively. The merger trees are derived by following the formation and merger history of each halo/subhalo with the D-Tree algorithm \citep{Jiang14}, based on which {\CODE} is applied to.

The simulation has a boxsize of $685\Mpc$ on a side, which is large enough compared to the detectable horizon of current ground based GW detectors, ranging from $58$ to $218\Mpc$ for NNMs \citep{Abbott17}. The mass resolution of dark matter particle in the Millennium Simulation is $1.2 \times10^9 \Msun$, allowing {\CODE} to generate a complete galaxy catalogue for galaxies more massive than $\sim10^9\Msun$.  

%In {\CODE}, the star formation history of each galaxy is stored in a linked-list during the galaxy assembly, i.e. whenever a star formation event happens a SSP is added into the galaxy, and a new node is attached to the linked-list simultaneously. Each node of the link-list records the initial mass, metallicity, formation site (disk or bugle) and formation time of a SSP. After tree running, we can calculate all kinds of star formation history related properties by using this linked-list, such as luminosity in different bands and $\RCOM$. 

\subsection{Stellar population synthesis models}
\label{section:SPSM}

{\YunnanII} stellar population synthesis model \citep{Zhang04,Zhang05,Zhang10} is used to model binary evolution in {\CODE}. {\YunnanII} is a stellar population synthesis model developed by the Group of Binary Population Synthesis of Yunnan Observatories. It is built based on the rapid binary star evolution (BSE) algorithm of \cite{Hurley02}, which modeled various binary interactions including mass transfer, mass accretion, common-envelope evolution, collisions, supernova kicks, tidal evolution and angular momentum loss through GWs. In \cite{Zhang10}, the evolutionary population synthesis models of \cite{Han07} which considered sub dwarf B stars (sdBs) are also included. With the help of {\YunnanII} model, instead of only modeling single star evolution as in \cite{BC03}, properties of binary stars in galaxies can be studied.

The setting of initial parameters of {\YunnanII} model and updated model parameters can be found in section 2.4.1 of \citealt{Jiang19}. Here we briefly describe the changes we have made in this work based on the fiducial {\YunnanII} model. 1) The range of the initial mass of the primary star in a binary is changed from $[0.1,100]\Msun$ to $[5,100]\Msun$, to focus on binaries that can have remnants of neutron stars and black holes that we are studying. 2) The initial mass function (IMF) of \citealt{Chabrier03} is used, replacing the approximated IMF given by \citealt{Eggleton89}. 3) The maximum mass of neutron star is set to be $3.0\Msun$, rather than $1.8\Msun$ in the oiriginal BSE. 4) The kick velocities of supernovae\footnote{The distribution of kick velocity of supernovae in stellar evolution model is generally fitted with a Maxwellian distribution. In this work, the parameter $\Vkick$ represents the dispersion of this Maxwellian distribution. $\Vkick$ applies to both the first and second supernovae during the formation of NNMs or NBMs. See Appendix A1 of \citealt{Hurley02} for more details.}, which are the natal velocities of the remnants after supernovae due to the asymmetry of explosion, are set to be able to vary in the range from $0\kms$ to $190\kms$, instead of the fixed value of $190\kms$ in the fiducial {\YunnanII} model. As we find that the value of kick velocity influence the merger rate a lot ($\sim2$ magnitude, see Fig. \ref{fig:DD} below). Large kick velocities will enlarge the orbital separations after supernova and delay the coalescence. Binary systems could even be tore apart with larger kick velocities. Thus increasing kick velocities will lower the merger rate. Besides, there are studies indicate that kick velocities in binary systems could be lower than ones of single stars in some cases (\citealt{Podsiadlowski04,Dewi05,Tauris15,Tauris17}; see section \ref{section:cosmic} for more details). Four values of $\Vkick=0,50,100,190\kms$ are applied and checked in section \ref{section:Rssp} and \ref{section:cosmic}. 
%Details of why and how we have made these changes compared to the fiducial {\YunnanII} can be found in Appendix A. 
From section \ref{section:rprocess}, $\Vkick=0\kms$ is chosen to build our fiducial model, since with this value the predicted cosmic event rate density is more consistent with the observational constraint of LVC in the local Universe, as can be seen in Fig. \ref{fig:NNmz} and section \ref{section:cosmic}.

Note that apart from the kick velocity of supernovae, other model parameters of stellar population synthesis model, such as common envelope parameter and mass transfer parameter, could also affect the merger rate \citep[e.g.][]{Dominik12,Chruslinska18}. While the focus of this work is the evolution of COMs and their host galaxies. A full exploration of the parameter space is beyond the scope of this paper. We leave this question in future works.

\subsection{Calculating event rates}
\label{section:rate}

For a simple stellar population (SSP)\footnote{A simple stellar population represents a set of stars formed together at the same time, having the same age and metallicity. ``Simple'' is used to be distinguished from the so-called complex stellar population, which is composed of multiple simple stellar populations.}, during the running of BSE algorithm as described in section \ref{section:SPSM}, we record every NNM and NBM event that occurs in the evolution process. By doing so, we get the compact object merger event rate $\RsspCOM(Z,\tage)$ for a SSP of certain age and metallicity.

%Input this initial condition library into the BSE algorithm, it then be transformed to a evolutionary track library. In order to derive $\RBNSMsp$, we track all these evolutionary tracks, whenever a BNSM happens, we write down an entry, including the age, metallicity and mass of each star of this BNSM. After tracking all the $10^7$ tracks, we statistic all the entries and tabulate them into $\RBNSMsp$, i.e. the distribution of all the events as the function of age and metallicity.

%As shown in the Fig. 1 of \cite{Metzger12}, the ejecta of kilonova is a rather isotropic structure and can be observed from a broad range of observer angle. In this work, we assume that kilonovae are spherical symmetric. 

%We use the most direct way to calculate $\RNNM$ and $\RNBM$ for each galaxy. 

For a galaxy that is comprised of millions to billions of stars with different mass, age and metallicity, the total COM rate in a galaxy at a certain time $t$, $\RCOM \left( t \right)$, can be calculated as the sum of the COM rate for all the SSPs in the galaxy:
\beq
\label{eq:RCOM}
\RCOM \left( t \right) = \int {\int_{0}^{t}{SFR \left( Z, \tau \right) \RsspCOM \left( Z, t-\tau \right) \dif \tau} \dif Z},
\eeq
where $SFR(Z,\tau)$ is the star formation rate of the galaxy at time $\tau$ for stars with metallicity $Z$, $\RsspCOM \left( Z, t-\tau \right)$ is the COM rate for a SSP of mass $1\Msun$ with metallicity $Z$ and age $t-\tau$. 
%In this work, $SFR$ is obtained from {\CODE} and $\RsspCOM \left( Z, t-\tau \right)$ is derived from {\YunnanII} SPSM.

In this work, we use the discretized version of Equ. (\ref{eq:RCOM}) to calculate $\RCOM$ in a galaxy at time $t$:
\beq
\label{eq:RCOM2}
\RCOM \left( t \right)  = \sum_{i=0}^{N_{\rm SP}}{M_{\rm i} \RsspCOM \left( Z_{\rm i},t-t_{\rm form,i} \right)},
\eeq
where $N_{\rm SP}$ is the total number of SSPs in this galaxy, and $M_{\rm i}$, $Z_{\rm i}$, $t_{\rm form,i}$ are the initial mass, metallicity and formation time of the $\rm i_{\rm th}$ SSP respectively. 

We assume that all NNMs and NBMs can produce kilonovae. In this case, the observed event rate of kilonovae $\Rkilonova$ can be written as:
\beqa
\Rkilonova &= &f_{\rm beam}(\RNNM+\RNBM)\nonumber \\
           &= &\RNNM+\RNBM = \RCOM.
\eeqa
$\RNNM$ and $\RNBM$ are the event rates of NNMs and NBMs respectively. $f_{\rm beam}$ is the beaming factor. As shown by \cite{Metzger12}, the ejecta of kilonova has a rather isotropic structure and can be observed from a broad angle range. Therefore we adopt $f_{\rm beam}=1$ in this work. 

Compared with NNM (see \citealt{Batiotti17} for a review), the mechanism and EM counterpart of NBM is much more ambiguous and still under debate. If the mass ratio of black hole over neutron star is very large, the neutron star will be swallowed into the black hole as a whole, and no EM emission is expected. Otherwise, the neutron star will be disrupted tidally beyond the Schwarzschild radius of the black hole and produce EM emission \citep{Shibata09}. Besides, even in the large mass ratio case, if the neutron star is highly magnetized or the black hole is charged, certain EM emission could be produced \citep{Mingarelli15,DOrazio16,Zhang19,Dai19}. On the observation side, several NBM candidates have been detected during the LVC O3. Among them, S190814bv is the most attracting one, as its false alarm rate is 1 per $1.559\times10^{25}$ years. However, no EM counterpart of S190814bv has been found so far \citep{Andreoni19,Gomez19,Dobie19}. The possible reasons may be that S190814bv is actually a BBM rather than a NBM, or the mass ratio of BH-NS is too large for EM emission as mentioned above, or the GW signal of S190814bv is a reflected one which arrives the Earth much latter than its EM signals \citep{Wei19}. In summary, we assume all NNMs and NBMs can produce kilonovae for simplicity. Which kind of NBMs can produce kilonovae and the fraction of them are still unclear, both theoretically and observationally.

\section{COM event rate and r-process production}
\label{section:results}

In this section, we show first the merger event rates of SSP for both NNMs and NBMs in the {\YunnanII} stellar population synthesis model. Then we study the cosmic COM rate density predicted in our {\CODE} semi-analytic model, by combining the $\RsspCOM$ of {\YunnanII} model with star formation histories using Equ. (\ref{eq:RCOM2}) for each galaxy. 
%The results presented include cosmic mean COM rate density, the properties of host galaxies of COMs, and r-process elements produced through COMs.
In section \ref{section:rprocess}, we present the amount of r-process elements produced by NNMs and NBMs in the lifetimes of galaxies. 

\subsection{Event rate in SSP}
\label{section:Rssp}

\begin{figure}
  \centering
  \includegraphics[width=0.5\textwidth]{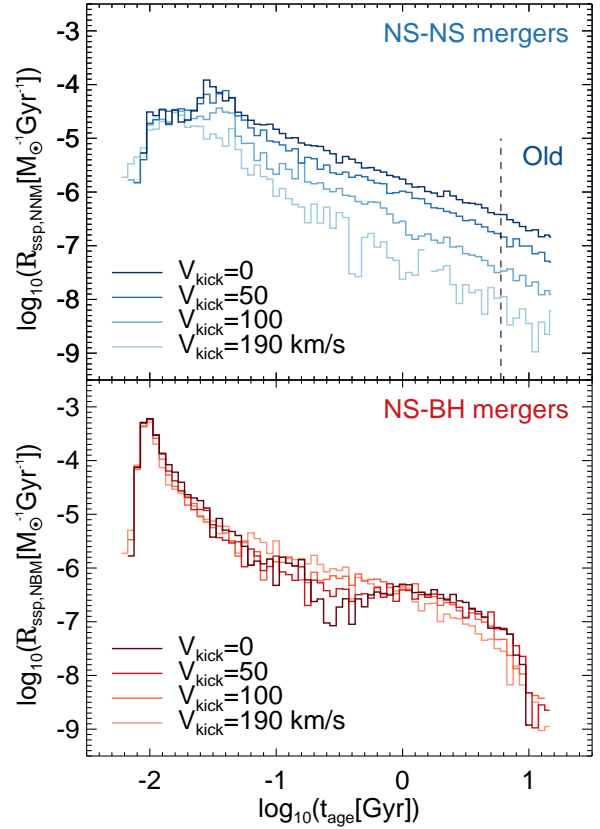}
  \caption{Event rates of NNM (upper panel) and NBM (lower panel) as a function of age for SSPs of $1\Msun$ with solar metallicity in {\YunnanII} model. Lines of different colors indicate results with different $\Vkick$ as shown in the label. The vertical dashed line in the upper panel marks the age of $6\Gyr$, which is the division of ``Old'' NNM population: NNMs with merger timescale longer than $6\Gyr$ is defined as Old NNMs.}
  \label{fig:DD}
\end{figure}

Fig. \ref{fig:DD} shows the event rates we derive from the {\YunnanII} stellar population synthesis model for SSPs, as a function of the age of the stellar population. $\RsspNNM$ and $\RsspNBM$ are event rates of NNMs and NBMs, and are presented in the upper and lower panel respectively. In each case, results for four different values of supernovae kick velocities are shown, and are all for solar metallicity.

In the upper panel of Fig. \ref{fig:DD}, for $\Vkick=0\kms$, we see that NNMs start to appear at $\sim 10\Myr$ after the birth of the stellar population. The event rate $\RsspNNM$ peaks at $\sim 30\Myr$ and then decreases as $\propto \tage^{-1}$, consistent with the theoretical expectation of the delay time distribution of COMs \citep{Maoz14,Toonen12,Yungelson13}.
%\footnote{If we assume the orbital separation $a$ when double compact object appears has the same distribution with their inital orbital separation, which is uniform in logarithmic space in {\YunnanII} model and most SPSMs: $\dif N/ \dif \log a \propto a^0$, i.e. $\dif N/\dif a \propto a^{-1}$, we have $\RsspCOM \propto \dif N/\dif t \propto t^{-1}$, as $t \propto a^4$ for GW dissipation. This is similar to the double degeneracy model of type Ia supernovae, which also losses angular momentum through GWs. \citep{Maoz14,Toonen12,Yungelson13}}. 
For different $\Vkick$, $\RsspNNM$ in general decreases as $\Vkick$ increases. The difference is small for $\tage < 20\Myr$, and can be as large as 2 dex at late times.
In the lower panel of Fig. \ref{fig:DD}, we see that at all ages the event rates are similar for different values of $\Vkick$, except that for $100\Myr < \tage < 1\Gyr$, $\RsspNBM$ with $\Vkick=0\kms$ is obviously lower than the one with $\Vkick=190\kms$. 

We have checked the model in detail and found two reasons responsible for the dependence of event rates on $\Vkick$ :
1) In {\YunnanII} model, the supernova which leaves a black hole as its remnant does not have kick velocity to the binary system, while the supernova which forms a neutron star has natal kick. %However, the orbital separation is still enlarged after this supernova because of mass loss. 
Therefore changing $\Vkick$ has smaller influence on $\RsspNBM$ than on $\RsspNNM$.
2) For $\Vkick=0\kms$, $\RsspNBM$ peaks at $\sim 10\Myr$ and also at $\sim 1\Gyr$. During the helium burning regime of the secondary star of a binary, if the star overfills the Roche-lobe, a common envelope forms. The orbital energy is then used to overcome the binding energy of common envelope, decreasing the separation tremendously. This NS-BH binary will coalesce in $\sim\Myr$, which corresponds to the peak at $\sim 10\Myr$. Otherwise, the secondary evolves to a neutron star independently and form a NS-BH binary with relatively large separation. The orbital energy is dissipated through GWs, and this NS-BH would coalesce in $\sim\Gyr$ scale. Increasing $\Vkick$ extends the time scales for mergers to happen, and makes the two peaks less distinct. 
%In one word, the natal kick in NS-BHs is weaker than the one in NS-NSs. Increasing $\Vkick$ only mildly modifies the distribution of merger timescale, rather than tears apart more and more binaries, as what happens for NS-NSs.

\begin{figure}
  \centering
  \includegraphics[width=0.5\textwidth]{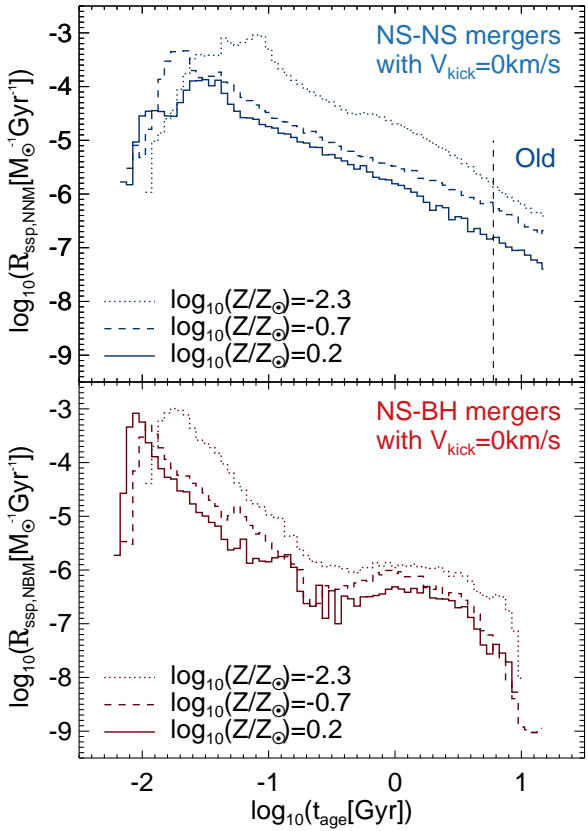}
  \caption{Event rates of NNM (upper panel) and NBM (lower panel) as a function of age for SSPs of $1\Msun$ with different metallicities in {\YunnanII} model with $\Vkick = 0\kms$. Lines of different line styles indicate results with different metallicities as shown in the label. The vertical dashed line in the upper panel marks the age of $6\Gyr$, which is the division of ``Old'' NNM population: NNMs with merger timescale longer than $6\Gyr$ is defined as Old NNMs.}
  \label{fig:DD_ageZ}
\end{figure}

In Fig.\ref{fig:DD_ageZ}, we show the dependence of $\RsspCOM$ on metallicity for $\Vkick = 0\kms$. We see that the event rates peak at earlier ages for higher metallcity, indicating shorter timescales of mergers to happen. 
%The merger timescale of NS-BH is relatively shorter than the one of NS-NS. $\RsspCOM$ has obvious evolution along the age of SSP (from $10^{-3}$ to $10^{-9}\Msun^{-1}\Gyr^{-1}$), 
Nevertheless, the dependence on metallicity for both COM rates is relatively weak (less than about a  magnitude), much less than the dependence on age.

%At last, we divide BNSMs into three populations: young ($<100\Myr$), mid-aged ($100\Myr - 6\Gyr$), and old population ($>6\Gyr$), as shown in the lower panel of Fig.(\ref{fig:DD}). Till now, more than a dozen BNSs have been observed as pulsars in binaries. There time remain to coalesce through gravitational waves can be estimated by their orbital parameters. Amoung them, PSR J0737-3039 has the shortest remain lifetime, which is $\sim 85\Myr$ (\cite{Burgay03}). However, there are many BNSs in {\YunnanII} model have coalescence time shorter than $50\Myr$, actually BNSM rate peaks at $30\Myr$. \cite{Gondek05} argue that short orbit BNSs are hard to observe, as their pulsar signal is blurred by orbital motion. Thus, the young population in our results represents such BNSs: exsit in stellar population synthesis models, but have not been observed. GW170817 is a direct detected BNS system by gravitational wave detectors, and its median merger timescale is $6.8-13.6\Gyr$ according to the age of its host galaxy NGC 4993 (\cite{Blanchard17}). The old population in our results represents BNSs like GW170817, which has long merger timescale comparable to the age of universe.

As mentioned in section \ref{section:intro}, the first directly detected NNM event is GW170817. The timescale from star formation to coalescence of this binary is larger than $6.8\Gyr$ with $90\%$ confidence according to the stellar mass build-up history of its host galaxy NGC 4993, which is inferred from its best-fit spectral energy distribution (SED) model \citep{Blanchard17}. To be compared with this specific observed event and check whether GW170817 in NGC 4993 is a typical NNM event, we study in particular the ``Old'' NNMs, defined as the NNMs that have survived longer than $6\Gyr$ before mergers happen, as shown in the upper panels of Fig. \ref{fig:DD} and Fig. \ref{fig:DD_ageZ}. From these panels we see that the ``Old'' NNMs that are like GW170817 are only a small fraction ($\sim 14\%$) of all NNMs.
For NBMs, the lower pannels of Fig. \ref{fig:DD} and Fig. \ref{fig:DD_ageZ} show that $\RsspNBM$ drops quickly in old SSPs ($>10\Gyr$), corresponding to few ``Old'' NBMs predicted in the model.

\subsection{Cosmic event rate density}
\label{section:cosmic}

\begin{figure}
  \centering
  \includegraphics[width=0.5\textwidth]{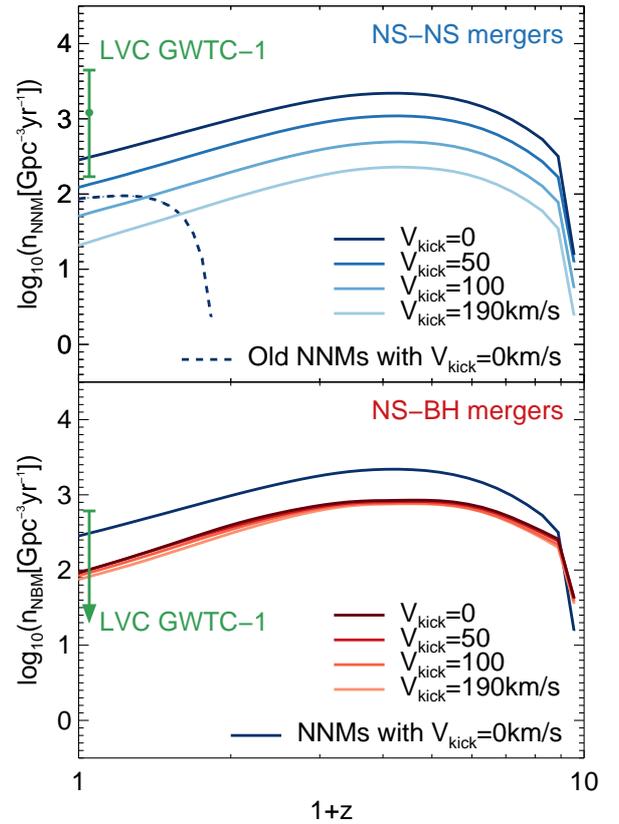}
  \caption{Upper panel: Cosmic NNM event rate per comoving volume as a function of redshift. Solid lines with different colors represent results with different $\Vkick$ as shown in the label. The dashed blue line shows the event rate density of Old NNMs with $\Vkick = 0$. The green dot with error bar is the observational result of LVC GWTC-1 at $z \sim 0$ \protect\citep{GWTC1}. Lower panel: The same as in the upper panel, but for NBMs. For comparison, the event rate density of NNMs with $\Vkick = 0$ is also shown as the blue solid line.}
  \label{fig:NNmz}
\end{figure}

In the previous subsection we show COM event rates in SSPs. From this subsection, we will use the semi-analytic model to predict the total event rate from the cosmological point of view, including  galaxies with different star formation histories comprised of complex stellar populations, using the method described in section \ref{section:rate}.

When accounting for all galaxies in the output of the semi-analytic model, Fig. \ref{fig:NNmz} gives the cosmic COM rate density as a function of redshift, where results from the NNMs and NBMs are shown in the upper and lower panels respectively. From the upper panel, we see that the NNM cosmic rate $\nNNM$ peaks at $z \sim 3.3$ and decreases gradually towards $z =0$, closely following the trend of the cosmic star formation rate density which peaks at $z\sim3.6$ in {\CODE}, with a short time delay in general ($\sim180\Myr$). The time delay agrees with the delay time distribution of NNMs as shown in Fig. \ref{fig:DD}. %most NNMs have short merger timescales compared with the age of the Universe. 
With different $\Vkick$, $\nNNM$ varies by a factor of $\sim 1$ dex. The number density with $\Vkick=0\kms$ at $z=0$ is $283 \Gpcyr$, marginally agree with the observational result of LVC GWTC-1 \citep{GWTC1}, which is $1210^{+3230}_{-1040}\Gpcyr$, as shown by the green dot with error bar in the upper panel of Fig. \ref{fig:NNmz}. Results with higher $\Vkick$ predict lower event rate densities, due to the generally lower event rate in SSP as shown in Fig. \ref{fig:DD}. Therefore, we choose $\Vkick=0\kms$, i.e. $\nNNM=283\Gpcyr$,  as our fiducial model, which our following analysis is based on. Compared with other works before, our fiducial model falls into their plausible ranges ($1.5-600\Gpcyr$, \citealt{Chruslinska18}; $20-600\Gpcyr$, \citealt{Mapelli18}; $238\Gpcyr$, \citealt{Artale19}).

%agreed with the theoretical expectation we mentioned above that kick velocities of binaries could be smaller than the one in single situation. 

%Results with different $\Vkick$ are also shown in Fig.(\ref{fig:NNmz}). We see that with higher $\Vkick$, $\rhoBNSM$ is lower. The difference can be as large as 1.5 dex at $z=0$. We give the prediction of $\rhoBNSM$ at $z=0$ is $\rhoBNSM = 619.8^{+3015.0}_{-497.6}\Gpcyr$. The observational density of LIGO O1 and O2 at $z \sim 0$: $1540^{+3200}_{-1220}\Gpcyr$ is shown as green error bar in Fig.(\ref{fig:NNmz}). Our prediction range agrees well with the observation. As the results with $\Vkick=50\kms$ fit the observation best, we treat it as the fiducial model in this work. We see that model with $\Vkick=190\kms$ which observed from single pulsars underestimates $\rhoBNSM$, too many binaries are torn apart. $\Vkick$ of tens of kilometre per second in binaries is needed in order to recover observed BNSM rate, agreed with theretical interpretation of the formation of BNSs (\cite{Tauris17}).

Blue dashed line in the upper panel of Fig. \ref{fig:NNmz} shows $\nNNM$ of Old NNMs (as defined in section \ref{section:Rssp}) with $\Vkick=0\kms$. Old NNMs start to appear at low redshift ($z<0.8$) when some stellar populations become older than $6\Gyr$, and have a larger fraction in all NNMs towards lower redshift. At $z=0$, Old NNMs is about $30\%$ of all NNMs. According to \cite{Blanchard17}, GW170817 is found in an old galaxy, and is probably an old NNM. While our result shows that younger NNMs have higher event rate density than Old NNMs in the local universe. More younger NNMs should be discovered (and in younger galaxies) with future observations, especially when GW detectors go to higher redshifts (the horizon of LIGO H1 during O2 is $218\Mpc$, i.e. $z<0.043$).

In the lower panel of Fig. \ref{fig:NNmz}, the evolution of the event rate density of NBMs, $\nNBM$, is shown. The trend is similar as for the evolution of $\nNNM$ shown in the upper panel, and also follows the trend of star formation rate density closely. The results do not vary much with different $\Vkick$, consistent with the results of SSPs as shown in the lower panel of Fig. \ref{fig:DD}. For $\Vkick=0\kms$, the NBM rate density is in general lower than the NNM rate density. At $z=0$, $\nNBM$ is $91\Gpcyr$, about a third of $\nNNM$, which is consistent with the upper limit proposed by LVC GWTC-1 shown as the green arrow.

\begin{table*}
  \caption{COM event rates in the model and in observation at $z=0$. For NNMs, our model results of the cosmic COM rate density $\nCOM$ for four different $\Vkick$ are listed. LVC GWTC-1 \protect\citep{GWTC1} result is also listed for comparison. For NBMs, only results with $\Vkick=0\kms$ are shown since the dependence of $\nNBM$ on $\Vkick$ is weak. The three right columns list the event numbers per year predicted for different GW detectors in each case. $\Vkick=0\kms$ is chosen as the fiducial model of this work and the related numbers are shown in bold. }
  \begin{tabular}{llcccc}
    \hline
     & $\Vkick$ & $\nCOM$ & & $N_{\rm COM}(\peryr)$ &\\
     & $(\kms)$ & $(\Gpcyr)$ & Virgo & LIGO L1 & LIGO H1\\
    \hline
    NNMs & $\mathbf{0}$ & $\mathbf{283}$ & $\mathbf{0.231}$ & $\mathbf{1.45}$ & $\mathbf{12.3}$\\
     & 50 & 123 & 0.100 & 0.629 & 5.32 \\
     & 100 & 51 & 0.041 & 0.26 & 2.2\\
     & 190 & 21 & 0.017 & 0.11 & 0.90\\
     & LVC GWTC-1 & $1210^{+3230}_{-1040}$ & $0.99^{+2.64}_{-0.85}$ & $6.21^{+16.57}_{-5.34}$ & $52.5^{+140.2}_{-45.1}$\\
    \hline
    NBMs & $\mathbf{0}$ & $\mathbf{91}$ & $\mathbf{0.075}$ & $\mathbf{0.47}$ & $\mathbf{4.0}$\\
    & LVC GWTC-1 & $<610$ & $<0.50$ & $<3.13$ & $<26.5$\\
    \hline
    \label{tab:nCOM}
  \end{tabular}
\end{table*}

The cosmic event rate densities at $z=0$ for NNMs and NBMs in different models are listed in Table. \ref{tab:nCOM}. We also show the event number per year that is predicted to be detected by Virgo and LIGO detectors, by assuming the detector horizons of NNMs and NBMs are $58\Mpc$, $107\Mpc$ and $218\Mpc$ for Virgo, LIGO L1 and LIGO H1 respectively \citep{Abbott17}.

Our models prefer a low kick velocity ($<50\kms$) for the progenitors of NNMs. Whereas, the observation of pulsar proper motions claim a much larger kick velocity: $190\kms$ \citep{Hansen97}, and $265\kms$ \citep{Hobbs05}. This inconsistency may shed some light on understanding the different formation routes of the isolated neutron stars and the ones in the pairs. Recent studies have proposed two formation channels for neutron stars with low kick velocity in binaries: 1) \cite{Podsiadlowski04} found that stars with initial mass $8-11\Msun$ in binary systems are likely to undergo an electron-capture supernova, rather than a neutrino-driven supernova as the case for a single star. Accretion induced collapse of massive white dwarfs with O/Ne/Mg cores can also lead to electron-capture supernovae. Electron-capture supernova is almost symmetric, short-duration and has smaller explosion energy \citep{Gessner18,Dessart06}, which naturally leads to an explosion with smaller kick velocity. 2) \cite{Tauris15,Tauris17} concluded that ultra-stripped supernovae (the second supernova in the formation of NS-NS and whose progenitor is an almost naked helium star) in close binaries generally have small kick velocities, due to the low mass ($<0.1\Msun$) and low binding energy of the helium envelope. However, these channels have relatively strict requirement for the initial parameters of binaries. For instance, electron-capture supernova only works for stars with initial mass $8-11\Msun$, and ultra-stripped supernova require the pre-supernova orbital period to be $1{\,\rm h}-2{\,\rm d}$. The fraction of ultra-stripped supernovae of all supernovae Ic is small ($<1\%$, \citealt{Tauris13}).

Note that LVC GWTC-1 only constrains the merged  NS-NSs, rather than the whole population. The initial conditions of such merged NS-NSs, as well as their evolutionary tracks and kick velocities, could be different from ones that not merged. It may be not appropriate to quantify kick velocities by using one parameter. A more refined and physical model for the determination of kick velocities may be needed. This is beyond the scope of this paper and we leave it to future works. More detection and better constraint from LVC O3 may also alleviate this inconsistency.

\subsection{r-process elements production}
\label{section:rprocess}

The ejected neutron-rich wind during a NNM or NBM provides an excellent environment for r-process nucleosynthesis, which is a nuclear process responsible for the production of about half of the elements heavier than iron \citep{Burbidge57,Cameron57,Meyer94}. The ejecta is extremely neutron-rich (with electron fraction $\sim0.05$), which allows nuclei to capture neutrons on a timescale faster than $\beta$-decay, and some neutron-rich isotopes can only be produced through r-process. r-process elements can be produced not only in NNMs and NBMs, but may also be produced in core collapse supernovae \citep{Wheeler98,Argast04,Arnould07} and high entropy winds from young neutron stars \citep{Woosley92}. 
%In Equ. (\ref{eq:Mrp}), $\sum{Others}$ represents other formation and reduction channels of r-process elements in galaxies except NNMs and NBMs. 
In this work, we focus on the r-process in NNMs and NBMs, and do not account for all other possible formation and reduction channels. 

%As we have recorded all NNMs and NBMs in the history of all galaxies, by assuming all these NNMs and NBMs have the same r-process production efficiency, 
We calculate the total r-process elements mass produced by NNMs and NBMs in a galaxy as:
\beq
\label{eq:Mrp}
\Mrp=M_{\rm ejecta,NNM}N_{\rm life,NNM}+M_{\rm ejecta,NBM}N_{\rm life,NBM},
\eeq
where $\NlifeNNM$ and $\NlifeNBM$ are the total number of NNMs and NBMs in the whole life of a galaxy, which are derived from {\CODE} directly.

$\MejectaNNM$ and $\MejectaNBM$ are the mass of produced r-process elements in one NNM and NBM event, or the ``yield'' of r-process elements. The uncertainties of the yields are huge. For NNMs with different neutron star masses, the ejecta masses can vary by a factor of 5, from $7.6\times10^{-3}\Msun$ to $3.9\times10^{-2}\Msun$, and can be larger or smaller than the ones of NBMs \citep{Korobkin12}. For NBMs with different black hole spin, the ejecta masses vary by a factor of $\sim200$ \citep{Bauswein14}. Besides, the equation of state of the neutron star, and detailed disk-ejecta configuration can also influence nucleosynthesis efficicy of NBMs by a factor of a few \citep{Tanaka14,Fernandez17}. Here in this work, for simplicity, we assume that NNMs and NBMs have the same yield, and are the same as the observational constrains of the ejacta mass of GW170817 \citep{Cote18}\footnote{Their Table. 1 and Table. 2 are compilations of various literatures \citep{Abbott17estimating,Arcavi17,Cowperthwaite17,Chornock17,Evans17,Kasen17,Kasliwal17,Nicholl17,Perego17,Rosswog17,Smartt17,Tanaka17,Tanvir17,Troja17}.}. The value we adopt is $\MejectaNNM=\MejectaNBM=0.01-0.04\Msun$, for r-process elements with $A>79$.

\begin{figure}
  \centering
  \includegraphics[width=0.5\textwidth]{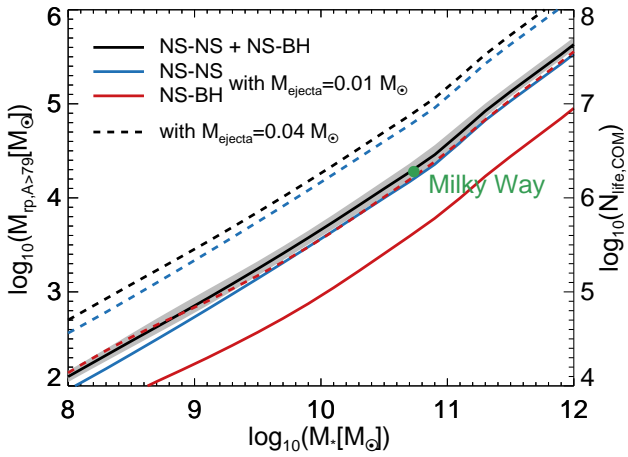}
  \caption{The mass of r-process elements with $A>79$ in a galaxy, $\MrpAs$, as a function of galaxy stellar mass at $z=0$ in our fiducial model. The blue and red solid lines are median $\MrpAs$ produced through NNM and NBM by assuming $\MejectaNNM=\MejectaNBM=0.01\Msun$, while the black solid line shows the sum of the two, with grey shadow indicates $1\sigma$ scatter around the median. The dashed blue, red and black lines are results when assuming $\MejectaNNM=\MejectaNBM=0.04\Msun$. The y axis on the right shows the corresponding number of COMs in the whole life of a galaxy, $N_{\rm life,COM}$, for the case of $\MejectaNNM=\MejectaNBM=0.01\Msun$. The green dot is the amount of observed r-process elements of the Milky Way: $M^{\rm obs}_{\rm rp,A>79,MW}=X_{\rm A>79}^{\rm obs}M_{\rm *,MW}=1.90^{+0.22}_{-0.21}\times10^{4}\Msun$, with $X_{\rm A>79}^{\rm obs}=35.0^{+0.4}_{-0.3}\times10^{-8}$ \protect\citep{Arnould07,Cote18} and $M_{\rm *,MW}=(5.43\pm{0.57})\times10^{10}\Msun$ \protect\citep{McMillan17}.}
  \label{fig:Mrp}
\end{figure}

Fig. \ref{fig:Mrp} gives the total mass of r-process elements with $A>79$ in a galaxy, $\MrpAs$, as a function of galaxy stellar mass at $z=0$ in our fiducial model. The solid and dashed lines are the predicted $\MrpAs$ by assuming $\MejectaNNM=\MejectaNBM=0.01\Msun$ and $\MejectaNNM=\MejectaNBM=0.04\Msun$ respectively, representing the lower and upper limits of our model prediction. 
%The blue and red lines are the $\MrpAs$ produced through NNMs and NBMs respectively, and the black lines are the sum of both channels. 
The $\Mstar-\MrpAs$ relation follows a power law with a scatter of only $\sim0.2$ dex, indicating that stellar mass determines the mass of r-process elements predominantly,
%The masses of r-process elements in galaxies with the same stellar mass only vary about $20\%$, even they have
much more than colors, star formation rates, metallicities, morphology, etc. The contributions from NNMs and NBMs are $\sim80\%$ and $\sim20\%$ at almost all stellar masses. 

By adopting the r-process mass fraction in the solar r-process residual, $X_{\rm A>79}^{\rm obs}=35.0^{+0.4}_{-0.3}\times10^{-8}$ \citep{Arnould07,Cote18}, and the Milky Way mass, $M_{\rm *,MW}=(5.43\pm{0.57})\times10^{10}\Msun$ \citep{McMillan17}, the observed mass of r-process elements in the Milky Way is $M^{\rm obs}_{\rm rp,A>79,MW}=X_{\rm A>79}^{\rm obs}M_{\rm *,MW}=1.90^{+0.22}_{-0.21}\times10^{4}\Msun$, as indicated by the green dot in Fig. \ref{fig:Mrp}. This observed value is in good agreement with our model prediction of $M_{\rm ejecta}=0.01\Msun$ , which is $M_{\rm rp,A>79,MW}=2.0^{+0.5}_{-0.4}\times10^{4}\Msun$ for Milky Way-mass galaxies (with $N_{\rm life,COM,MW}=2.0^{+0.5}_{-0.4}\times10^{6}$). The model of $M_{\rm ejecta}=0.04\Msun$ (dashed lines) overestimates the amount of r-process elements in Milky Way-mass galaxies. If the yield is indeed $0.04\Msun$, NBMs alone can provide sufficient amount of r-process elements in the Milky Way (the red dashed line).

\begin{figure}
  \centering
  \includegraphics[width=0.5\textwidth]{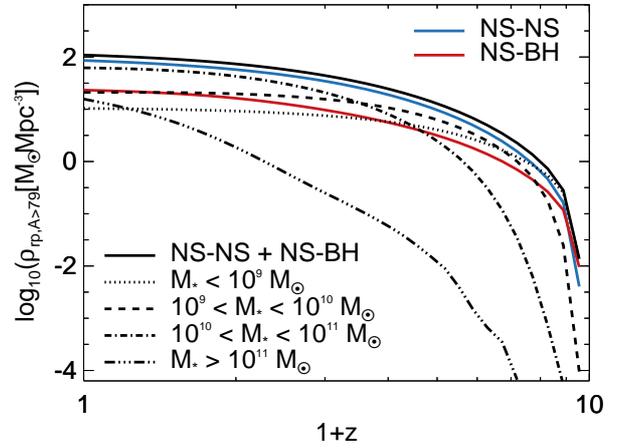}
  \caption{Cosmic amount of r-process elements with $A>79$ in galaxies per comoving volume, $\rho_{\rm rp,A>79}$, as a function of redshift in our fiducial model, and with $\MejectaNNM=\MejectaNBM=0.01\Msun$. The blue and red solid lines are the results of NNMs and NBMs respectively, and the black solid line is the sum of both channels. Black lines with different line styles are the amount of r-process elements contributed by galaxies with different stellar mass ranges, as shown in the label.}
  \label{fig:zrp}
\end{figure}

In Fig. \ref{fig:zrp}, we show the cosmic density evolution of the amount of r-process elements with $A>79$ predicted in our model. Only the model with $\MejectaNNM=\MejectaNBM=0.01\Msun$ is shown here, considering that adopting a different yield would not change the general trend of the result. As stellar mass is a good indicator of r-process elements as seen in Fig. \ref{fig:Mrp}, the amount of r-process elements accumulated gradually as the universe evolves towards low redshift. About $50\%$ of r-process elements nowadays already existed at $z\sim1.6$, and about $90\%$ r-process elements nowadays formed before $z\sim0.3$. The contributions from NNMs and NBMs are always $\sim80\%$ and $\sim20\%$ respectively, except for the earliest redshifts, due to the fact that NBMs have higher event rate than NNMs in young stellar populations (as shown in Fig. \ref{fig:DD}). We also explore the amount of r-process elements in galaxies with different stellar masses, as shown by black lines with different line styles. At $z=0$, most r-process elements ($\sim57\%$) are stored in galaxies with $10<\log{(\Mstar[\Msun])}<11$, which means Milky Way-mass galaxies are the main sites for historical r-process nucleosynthesis.

Our fiducial model ($\nNNM=283\Gpcyr$ and $\nNBM=91\Gpcyr$) with $\Meject=0.01\Msun$ matches the observed abundance of the Milky Way very well. Note that we have only calculated the r-process nucleosynthesis through NNMs and NBMs in the model. On the other hand, heavy elements abundances are usually measured through meteorite, solar spectra, and stellar spectra  \cite[e.g.][]{Anders89,Kappeler89}, which are all in stellar component. However, a substantial fraction of heavy elements could stay in gas phase, which may cause the underestimation of heavy elements production efficiency in current observation. Besides, due to the spread delay time distribution of NNMs, the lifetime of some NS-NSs can be comparable with the age of the Universe. Thus the offset between the location of coalescence and star-forming region can be $\ge 20\kpc$ \citep{Fong13}, which would also lower the amount of observed heavy elements leaving in galaxies.
%\textcolor{blue}{Therefore, although our fiducial model ($\nNNM=283\Gpcyr$ and $\nNBM=91\Gpcyr$) with $\Meject=0.01\Msun$ matches the observed abundance of the Milky Way very well, we are not able to draw robust conclusion that the model is indeed consistent with observation, before we build models in more detail to be more comparable with observation, and include contributions from all possible formation and reduction channels.}
%we could only conclude that NNMs alone or NNMs plus NBMs can provide the amount of r-process elements we observe, and the $0.04\Msun$ upper limit of the ejecta mass observed in GW170817 may provide even more than we need. Stronger conclusion can only be made after more thorough consideration, including other formation and reduction channels.

\section{Properties of COM Host Galaxy}
\label{section:host}
\subsection{Stellar mass and age}
\label{section:hostmass}

\begin{figure}
  \centering
  \includegraphics[width=0.5\textwidth]{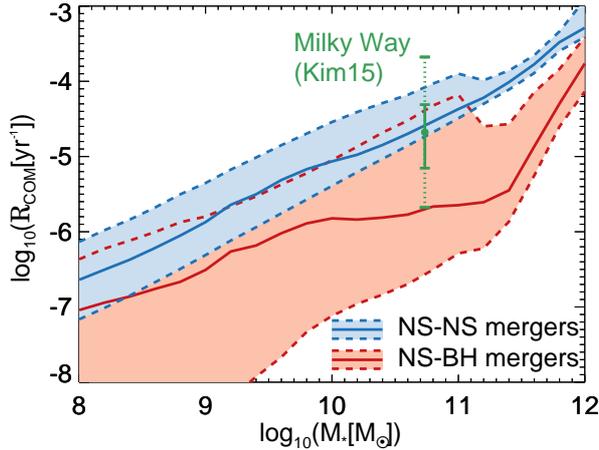}
  \caption{The total event rates of NNMs and NBMs in a galaxy, as a function of galaxy stellar mass at $z=0$ in our ficucial model. The solid lines show the median relations (blue for NNMs and red for NBMs). The light color regions and dashed lines indicate the $1\sigma$ scatter. The green dot with solid error bars is the observational result of \protect\cite{Kim15} for the NNM rate in the Milky Way, and the dotted error bars present the uncertainty after considering different assumptions about the pulsar luminosity distribution \protect\citep{Chruslinska18}.}
  \label{fig:RCOM}
\end{figure}

Fig. \ref{fig:RCOM} shows the total event rates of NNMs and NBMs in a galaxy as a function of galaxy stellar mass at $z=0$ in our fiducial model with $\Vkick=0\kms$.  The median of the $\Mstar-\RNNM$ relation can be well fitted by a power law, while $\Mstar-\RNBM$ deviates from a single power law and has larger scatter. Compared with NNMs, there are more NBMs in young SSPs ($\tage < 20\Myr$, as shown in Fig. \ref{fig:DD}). Therefore $\RNBM$ is more sensitive to recent star formation activities, which results in larger scatter. 

The observational result of \cite{Kim15} for the NNM rate in the Milky Way ($21^{+28}_{-14}\Myr^{-1}$) is shown as the green dot with error bars in Fig. \ref{fig:RCOM}. At the Milky Way mass of $M_{\rm *,MW}=5.43\times10^{10}\Msun$ \citep{McMillan17}, the predicted $\RNNM$ in our model is $R_{\rm NNM,MW}=25.7^{+59.6}_{-7.1}\Myr^{-1}$, in good agreement with the observation. The predicted $\RNBM$ for Milky Way-like galaxies is much lower, with $R_{\rm NBM,MW}=2.1^{+40.3}_{-1.8}\Myr^{-1}$.

\begin{figure*}
  \centering
  \includegraphics[width=1.0\textwidth]{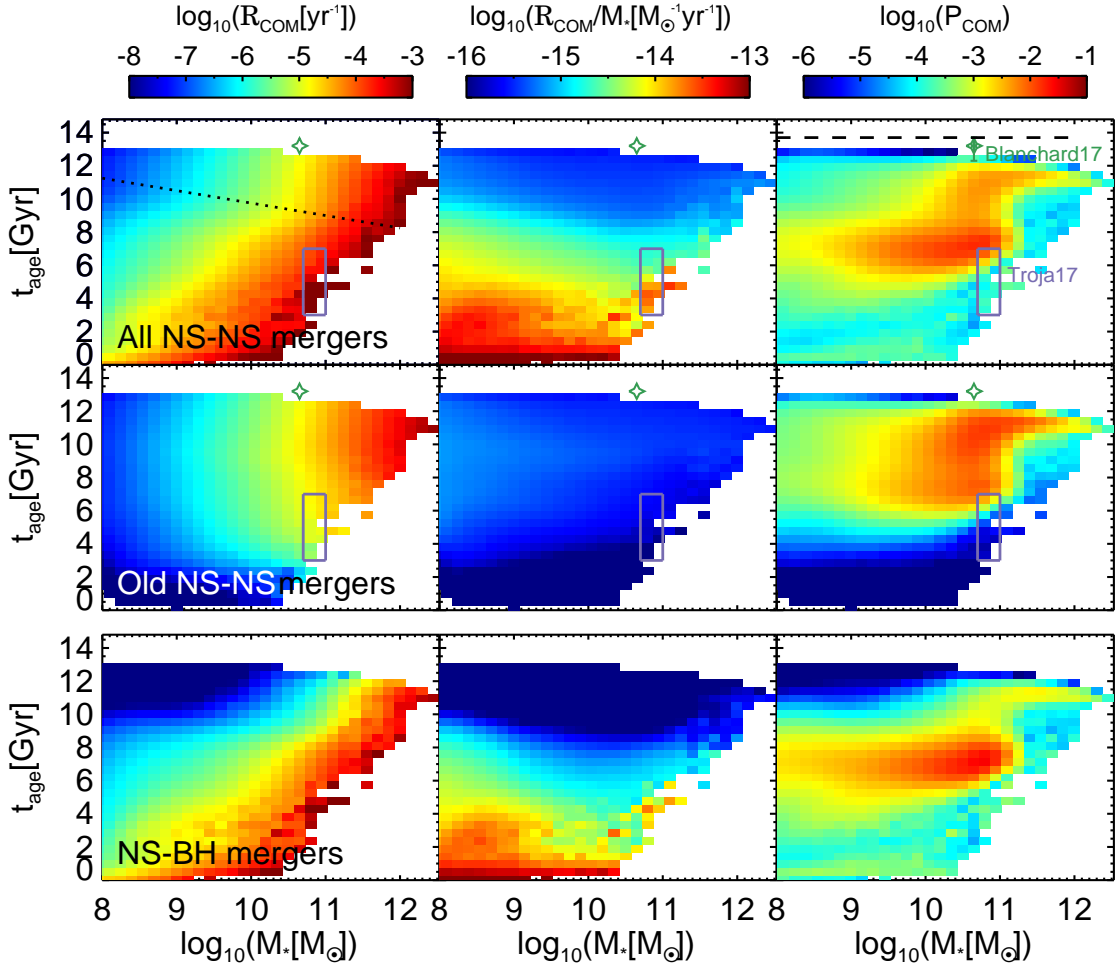}
  \caption{Distributions of event rate as a function of galaxy stellar mass and age at $z=0$ in our fiducial model. Columns from left to right show results of: the COM rate, COM rate per stellar mass, and $\PCOM$ (the probability to observe a merger event in a certain kind of galaxy as defined by Equ. (\ref{eq:PCOM})). The upper panels and middle panels are distributions of all NNMs and Old NNMs respectively, and the bottom panels are results of NBMs. In each panel, colors show the mean value for galaxies in each cell as indicated in the color bars. The dotted line in the upper left panel is $\tage[\Gyr]=-0.75*\log{(\Mstar[\Msun])}+17.25$, which divides the distribution into two populations. The green star represents NGC 4993, the host galaxy of GW170817, with $\Mstar=4.47\times10^{10}\Msun$ and $\tage=13.2\Gyr$ \protect\citep{Blanchard17}, with error bar only shown in the upper right panel. The purple rectangle also represents NGC 4993, but is the observational result of \protect\cite{Troja17}, with $\Mstar=(5-10)\times10^{10}\Msun$ and $\tage=3-7\Gyr$.  The dashed line in the upper right panel is $\tage=13.7\Gyr$, representing the age of the Universe. }
  \label{fig:age}
\end{figure*}

As seen in Fig. \ref{fig:DD_ageZ}, age influences COM event rate the most for SSP, so we check also the dependence of $\RCOM$ on the age of host galaxies\footnote{The age of a galaxy in this work is the so-called ``mass weighted age'', i.e. the mean value of ages of all the simple stellar populations, weighted by the initial mass of each stellar population.}. In the left column of Fig. \ref{fig:age}, the distributions of COM event rates in galaxies at $z=0$ of our fiducial model are plotted, in the stellar mass -- galaxy age plane. The upper left panel is the result of $\RNNM$, which shows that massive/young galaxies generally have higher $\RNNM$ than small/old galaxies, consistent with expectation as young stellar populations have higher event rate. The distribution can be divided into two populations as shown by the dotted line: for old galaxies above the line, there is almost no dependence on age; for galaxies younger, there exists clear dependence on both stellar mass and age.

The distribution of event rates for Old NNMs is shown in the middle left panel of Fig. \ref{fig:age}. For the youngest galaxies with $\tage<2\Gyr$, the event rates of Old NNMs are always low, independent of galaxy stellar mass. For galaxies with $\tage>2\Gyr$, $\RNNM$ of Old NNMs have strong dependence on stellar mass but almost no dependence on age. Compared with the case for all NNMs, the old galaxies above the dotted line in the upper left panel 
%the amount of Old NNMs (or Old SSPs with $\tage > 6\Gyr$) in a young galaxy or an old galaxy is roughly the same, as long as they have the same stellar mass. These Old SSPs have been distributed quite evenly in different kinds of galaxies through galaxy mergers. The $\tage>9\Gyr$ population in the upper left panel of Fig. \ref{fig:age} 
is mainly contributed by Old NNMs.
%, and have event rates with little dependence on age.
The distribution of $\RNBM$ is shown in the lower left panel of Fig. \ref{fig:age}, which has similar trend as that of $\RNNM$, except that without Old NBMs as shown in section \ref{section:Rssp}, the distribution of $\RNBM$ always depends both on age and stellar mass.

In the middle column of Fig. \ref{fig:age}, we show the distribution of specific event rate defined as $\RCOM$ divided by stellar mass. The dependence on stellar mass is largely reduced in this case.
%, where the $\RNNM$ of all NNMs almost only depend on the age of host galaxies, and the $\RNNM$ of Old NNMs also depend on the age, but the dependence is much weaker. 
%The $\RNNM$ per stellar mass actually represents the $\RNNM$ of a SSP to some degree, thus these distributions are the ``macro'' version of Fig. \ref{fig:DD_ageZ}, after considering the formation and evolution of galaxies. 
For both NNMs and NBMs, the specific event rate is slightly higher for less massive galaxies at given galaxy age. This is because low mass galaxies tend to have lower metallicity, while low metallicity results in relatively higher $\RsspCOM$ as shown in Fig.\ref{fig:DD_ageZ}.

In order to compare our results with observation, we need to figure out the probability to observe a COM event in a certain kind of galaxies, considering the number density of galaxies into account. We calculate the probability to detect a certain kind of galaxies as the host galaxy of a COM event as:
\beq
\label{eq:PCOM}
P_{\rm COM,i} = \frac{R_{\rm COM,i}N_{\rm gal,i}}{\sum_{i}{R_{\rm COM,i}N_{\rm gal,i}}},
\eeq
where the subscript $i$ stands for a certain kind of galaxies. $R_{\rm COM,i}$ is the mean COM rate and $N_{\rm gal,i}$ is the number density of this kind of galaxies. For example, the number densities of galaxies with certain stellar mass and age are presented in Fig. \ref{fig:agedis} for {\CODE} (see Appendix \ref{section:numberdensity} for details).
Note that in this work we do not consider the selection effect in observations, and assume all the host galaxies of COMs can be observed, which may overestimate the number of small host galaxies. The influence of selection effect will be explored in future works.

%Whether we can easily observe BNSM events in a certain kind of galaxies, not only determined by BNSM rate, but also influenced by the number density of that kind of galaxies. From above we know that young massive galaxies have the largest BNSM rates, but the number density of such galaxies is definitely low. The probability to detect a certain kind of galaxies as the host galaxy of a BNSM event can be calculated simply by:
%\beq
%\label{eq:PBNSM}
%P_{\rm BNSM,i} = \frac{R_{\rm BNSM,i}N_{\rm gal,i}}{\sum_{i}{R_{\rm BNSM,i}N_{\rm gal,i}}},
%\eeq
%where the subscript $i$ represents for a certain kind of galaxies, e.g. can be a cell in Fig.(\ref{fig:age}); $N_{\rm gal,i}$ is the number density of that kind of galaxies. Fig.(\ref{fig:agedis}) is the distribution of $\Ngal$. We see that small galaxies have much higher numbers than massive galaxies, and the peak is around $8\Gyr$. The upper right panel in Fig.(\ref{fig:age}) is the distribution of $\PBNSM$ for all BNSMs. Compared with the upper left panel, $\PBNSM$ has the similar trend as $\RBNSM$: massive and young galaxies are more likely detected as host galaxies, nevertheless, there is a another trend towards the $8\Gyr$ number density peak of small galaxies, making mid-aged small galaxies have higher detection rate then young small galaxies, even the latter ones have higher BNSM rate.

The predicted distributions of $\PCOM$ as a function of stellar mass and age are shown in the right column of Fig. \ref{fig:age}. We find that galaxies with $\Mstar=10^{10.65}\Msun$ and $\tage=7.1\Gyr$ are most likely detected as the host of a NNM event, as well as of a NBM. In the upper right panel of Fig. \ref{fig:age}, compared with the observational results of NGC 4993, the host galaxy of GW170817 (green star: \citealt{Blanchard17}, hereafter Blacnchard17; purple rectangle: \citealt{Troja17}, hereafter Troja17), we see that the peak of our prediction is marginally consistent with Troja17. Blanchard17 gives similar stellar mass, but the age is much older (close to the age of the Universe, as represented by the horizontal dashed line). 

%$\PBNSM$ of old BNSMs is also similar to its $\RBNSM$ distribution: old BNSMs are tend to be discoved in massive galaxies, no matter whether they are young or not. The inclusion of number density increases the detection probability of small galaxies with $age \sim 8\Gyr$.

\begin{figure}
  \centering
  \includegraphics[width=0.5\textwidth]{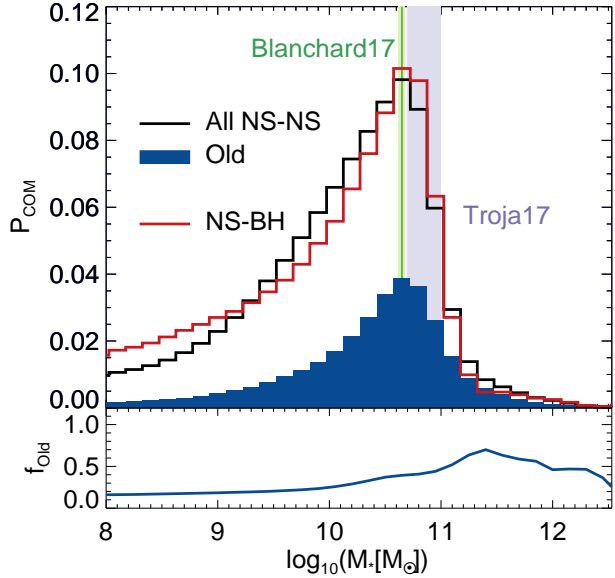}
  \caption{Upper panel: the stellar mass distribution of host galaxies of NNMs (black solid line) and NBMs (red solid line) at $z=0$ in our fiducial model. The blue area gives the contribution of Old NNMs. The green and purple regions represent the observed stellar mass of NGC 4993 by  \protect\cite{Blanchard17} and \protect\cite{Troja17}, respectively. Lower panel: the fraction of Old NNMs over all NNMs as a function of stellar mass: $\fOld=P_{\rm NNM,Old}/\PNNM$.}
  \label{fig:age_Px}
\end{figure}

\begin{figure}
  \centering
  \includegraphics[width=0.5\textwidth]{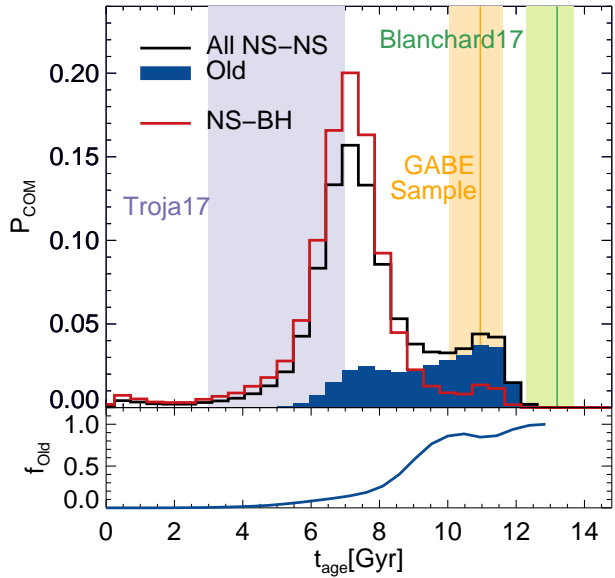}
  \caption{Upper panel: the age distribution of host galaxies of NNMs (black solid line) and NBMs (red solid line). The blue area is the contribution of Old NNMs. The green and purple regions represent the observed age of NGC 4993 provided by \protect\cite{Blanchard17} and \protect\cite{Troja17}, respectively.  The orange line and area are the predicted median and $1\sigma$ scatter of the age of selected NGC 4993-like galaxies in {\CODE} catalogue, which is described in detail in section \ref{sec:other}. Lower panel: the fraction of Old NNMs over all NNMs as a function of galaxy age.}
  \label{fig:age_Py}
\end{figure}

By integrating $\PCOM$ along $\tage$ ($\Mstar$) , we can derive the stellar mass (age) distribution of the host galaxy of COMs, and the results are shown in Fig. \ref{fig:age_Px} (Fig. \ref{fig:age_Py}). From Fig. \ref{fig:age_Px}, we see that the distributions of host galaxies of both NNMs and NBMs peak at around $10^{10.65}\Msun$, decreasing fast towards low mass end and even faster towards high mass end. The contribution of Old NNMs is about $\sim20\%$ for galaxies with $\Mstar < 10^{10}\Msun$. For more massive galaxies, Old NNMs contribute more, with a fraction as high as $\sim70\%$. The observational results of Blanchard17 and Troja17 of NGC 4993 are shown by green and purple shadow in Fig. \ref{fig:age_Px}, both lying around the peak of model prediction.

As presented in Fig. \ref{fig:age_Py}, the age distribution of host galaxies of NNMs is bimodal, with two peaks of around $\tage=7.1\Gyr$ and $\tage=11.0\Gyr$. The latter is mainly contributed by Old NNMs. Observationally, the age provided by Blanchard17 and Troja17 differs a lot from each other, which reflects the huge systematic uncertainties in determining age through galaxy spectrum. Unlike the age distribution of the host galaxies of NNMs, the one of NBMs only has the young peak because there is few Old NBMs in the model.

%Integral $\PBNSM(\Mstar,age)$ along $\Mstar$ or $age$, we get the stellar mass or age distribution of BNSMs' host galaxies, as shown in Fig.(\ref{fig:age_Px}) and Fig.(\ref{fig:age_Py}). From Fig.(\ref{fig:age_Px}), we see that the stellar mass distribution of BNSMs' host galaxies has a peak around $10^{10.68}\Msun$, decreasing fast towards low mass end, and even faster towards high mass end. The old BNSMs have the similar distribution with all BNSMs, which also peak around $10^{10.68}\Msun$. The contribution of old BNSMs in each bin is stable for galaxies $\Mstar < 10^{11}\Msun$, about $25\%$; for more massive galaxies, old BNSMs contribute more, can up to $70\%$. From Fig.(\ref{fig:age_Py}), we see that the age distribution of host galaxies peaks around $7.3\Gyr$, and there is another peak around $10.5\Gyr$, which is four times lower than the younger peak and mainly contributed by old BNSMs. The contribution of old BNSMs is small for galaxies $< 8\Gyr$, and increases fast as age getting oldder.

The main results of this subsection is that young and massive galaxies have higher COM rate. 
%The mass dependence of COM rate exists for all COMs, and the 
The age dependence is mainly caused by young COMs. Considering number densities of galaxies into account, COMs are most likely to be observed in galaxies with $\Mstar \sim 10^{10.65}\Msun$ and $\tage \sim 7.1\Gyr$, $11.0\Gyr$.

%The star in Fig.(\ref{fig:age}) and the green area in Fig.(\ref{fig:age_Px}), Fig.(\ref{fig:age_Py}) is NGC 4993, the host galaxy of GW170817 (\cite{Blanchard17}). It is an extremely old galaxy, with the mass-weighted age from spectrum analysis of $13.2^{+0.5}_{-0.9}\Gyr$. Its stellar mass is $10^{10.65^{+0.03}_{-0.03}}\Msun$. (NGC 4993 is out of the range of our model's coverage, this is because ?) From Fig.(\ref{fig:age}), we see that NGC 4993 has relatively high BNSM rate, and mostly contributed from old population BNSMs. It is not strange to discover the first BNSM in a galaxy like NGC 4993. In Fig.(\ref{fig:age_Px}), NGC 4993 locates right at the peak of stellar mass distrition, agrees extremely well with theoretical expectation. However, in Fig.(\ref{fig:age_Py}), NGC 4993 is outside of our model's coverage. The number of galaxies older than $11.5\Gyr$ declines greatly in our model, and there is no galaxies older than $13\Gyr$. NGC 4993 is beyound this limit, which is too old for our model. (?)For a stellar mass selected sample, NGC 4993 is a typical host galaxy of BNSM, we expect more galaxies around $10^{10.65}\Msun$ to be discoved as host galaxies in the future.

\subsection{Colors, sSFR, metallicity and morphology}
\label{sec:other}

\begin{figure*}
  \centering
  \includegraphics[width=1.0\textwidth]{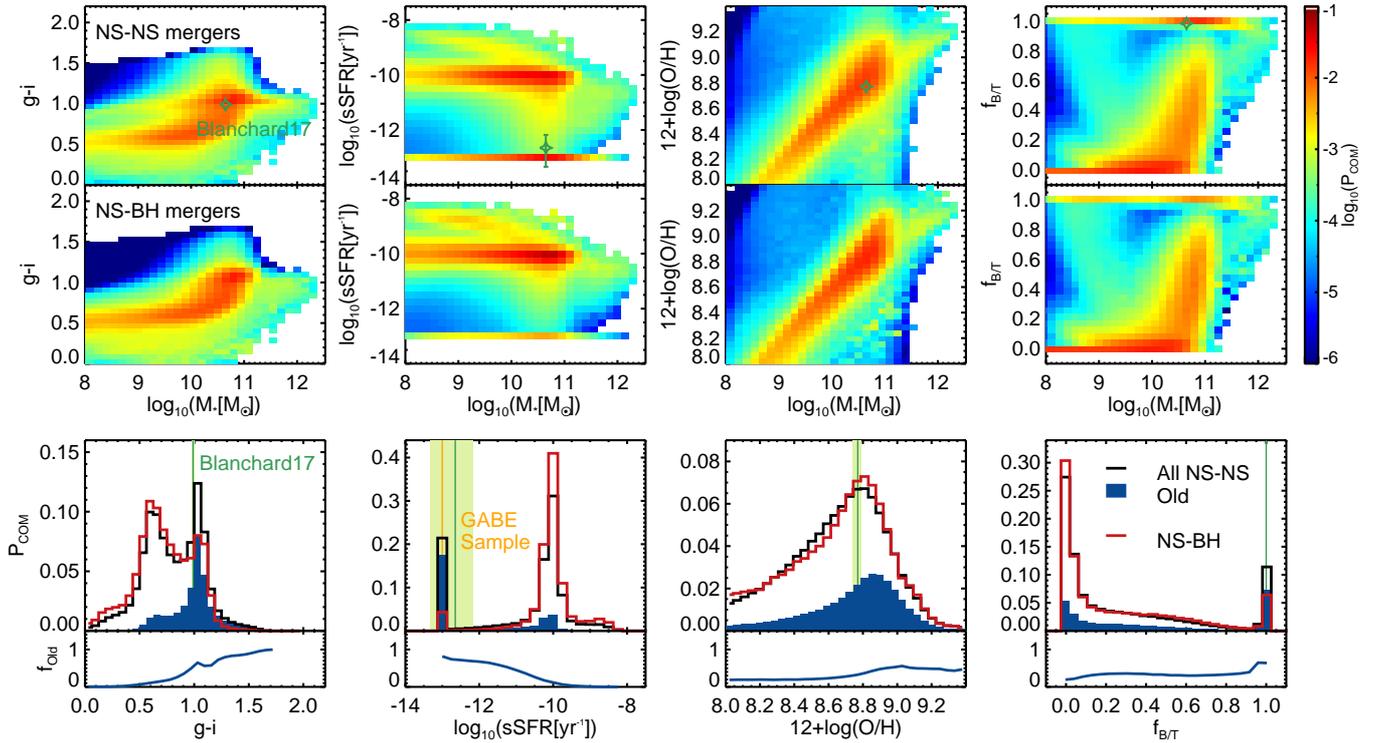}
  \caption{Panels from left to right: the distributions of $\PCOM$ (the probability to observe a merger event in a certain kind of galaxy as defined by Equ. (\ref{eq:PCOM})) as a function of g-i color, sSFR, gas phase metallicity, bulge-to-total stellar mass ratio ($\fBT$) of COM host galaxies and their stellar mass. The upper and middle rows are the results of NNMs and NBMs. Colors show the mean value in each cell as indicated in the color bar. The green stars with error bar represent the observational result of NGC 4993 in \protect\cite{Blanchard17}, as listed in Tab. \ref{tab:NGC4993}. The lower panels are the g-i color, sSFR, gas phase metallicity and $\fBT$ distributions of COM host galaxies, which are derived by integrating the distributions in the upper panels along stellar mass. The black and red line are the distributions for NNMs and NBMs respectively. Blue region is the contribution of Old NNMs. The green area represent the observational result of NGC 4993 of \protect\cite{Blanchard17}, the same as the star in upper panels. The orange line is the predicted median sSFR of selected NGC 4993-like galaxies in {\CODE} catalogue.}
  \label{fig:other}
\end{figure*}

Following the analysis on galaxy stellar mass and age as shown in Fig. \ref{fig:age}-\ref{fig:age_Py}, in this subsection, we further explore the dependence of $\PCOM$ on color, specific star formation rate (sSFR)\footnote{In this work, for model galaxies with sSFR lower than $10^{-13}\peryr$, we set their sSFR to be $10^{-13}\peryr$, in order to make comparison with observations.}, gas phase metallicity and bulge-to-total stellar mass ratio ($\fBT=M_{\rm *,bulge}/(M_{\rm *,bulge}+M_{\rm *,disk})$) for host galaxies of COMs. The results are presented in Fig. \ref{fig:other}.
The upper two rows of Fig. \ref{fig:other} show the distributions of $\PNNM$ and $\PNBM$ respectively, on the planes of galaxy properties and galaxy stellar mass. The host galaxies of NNMs and NBMs have similar distributions for all the galaxy properties investigated. 

%The upper panels are the distribution of $\PBNSM$. We see that the trend is similar with Fig.(\ref{fig:age}): for galaxies $\Mstar > 10^{10}\Msun$, no matter what other property is, $\PBNSM$ is relatively high, stellar mass is the dominant factore; for galaxies $\Mstar < 10^{10}\Msun$, blue, star-forming or metal-poor galaxies have higher BNSM rate, which is because these galaxies are younger.

%\begin{figure*}
%  \centering
%  \includegraphics[width=1.0\textwidth]{./figs/NNmMstar_fig12_1.ps}
%  \caption{panels from left to right shows the g-i color, sSFR, gas phase metallicity and bulge-to-total stellar mass ratio distribution of host galaxies. $\PBNSM$ is the ratio of the BNSMs number in each bin over total number of BNSMs, and the binsize is $0.0625$, $0.375\peryr$, $0.04375$, and $0.03125$ respectively. The light blue and the dark blue area is the contribution of young and old BNSMs in each age bin respectively. The green area represents the observation of NGC 4993: $g-i=0.99^{+0.01}_{-0.01}$, $sSFR=-12.65^{+0.47}_{-0.69}\peryr$, and $12+\log{\rm (O/H)}=8.77^{+0.02}_{-0.02}$, derived from \protect\cite{Blanchard17}. The lower panel of each panel: the contribution of old BNSMs.}
%  \label{fig:other2}
%\end{figure*}

By Integrating $\PCOM$ along stellar mass, we can derive the distributions of color, sSFR, metallicity and morphology for COM host galaxies in our model, as shown in the lower panels of Fig. \ref{fig:other}. The distributions of color, sSFR and morphology have two peaks, a red/quiescent/early-type sequence, and a blue/star-forming/late-type sequence that includes more galaxies, for both NNMs and NBMs. {\it Therefore it is more likely to detect COMs in blue/star-forming/late-type galaxies.} As expected, Old NNMs contribute more in red/quiescent/early-type sequence of all NNMs. The gas metallicity has a wide distribution and peaks at around $8.72-8.85$, which is comparable to solar metallicity (the solar metallicity in unit of $12+\log{\rm (O/H)}$ is $8.7$). Old NNMs contribute more in metal rich NNMs galaxies.

\begin{table}
  \caption{Properties of NGC 4993. Observational results given by \protect\cite{Blanchard17} and \protect\cite{Troja17} are listed, including stellar mass, mass-weighted age from spectrum analysis, g-i color, specific star formation rate (sSFR) and gas phase metallicity in unit of $12+\log{\rm (O/H)}$ (transferred from the [Fe/H] in Blanchard17). The last column is NGC 4993-like galaxies selected from {\CODE} galaxy catalogue, according to galaxy stellar mass, g-i color and metallicity as listed in this table. The predicted age and sSFR of selected NGC 4993-like galaxies are also listed and marked in bold.}
  \begin{tabular}{lcccp{0.5\textwidth}}
    \hline
    Property & Blanchard17 & Troja17 & GABE Sample\\
    \hline
    $\log{(\Mstar/\Msun)}$ & $10.65^{+0.03}_{-0.03}$ & $10.7-11.0$ & $10.5-10.8$\\
    age($\Gyr$) & $13.2^{+0.5}_{-0.9}$ & $3-7$ & $\mathbf{10.95^{+0.68}_{-0.90}}$\\
    g-i & $0.99^{+0.01}_{-0.01}$ & $-$ & $0.95-1.05$\\
    $\log{\rm (sSFR/\peryr)}$ & $-12.65^{+0.47}_{-0.69}$ & $-$ & $\mathbf{-13}$\\
    $12+\log{\rm (O/H)}$ & $8.77^{+0.02}_{-0.02}$ & $-$ & $8.7-8.9$\\
    \hline
    \label{tab:NGC4993}
  \end{tabular}
\end{table}

For NGC 4993 that is observed to host GW170817, we list in Table. \ref{tab:NGC4993} its properties derived by Blanchard17 and Troja17, and over-plot the values from Blanchard17 in Fig. \ref{fig:other} to be compared with model predictions. The observed values are always at or close to the peaks of the model distributions.

As listed in Table. \ref{tab:NGC4993}, the observed stellar mass, color and metallicity of NGC 4993 have small errors, much smaller than that of age and sSFR. Based on these ``accurate'' properties, we define elliptical galaxies ($\fBT>0.9$) with $\Mstar = 10^{10.5}-10^{10.8}\Msun$, $g-i=0.95-1.05$ and $12+\log{\rm (O/H)} = 8.7-8.9$ as NGC 4993-like galaxies. We find 7604 such galaxies from {\CODE} galaxy catalogue at $z=0$, and construct a NGC 4993-like galaxies sample. Their median age and $1\sigma$ scatter is $10.95^{+0.68}_{-0.90}\Gyr$, as shown by the orange region in Fig. \ref{fig:age_Py}, which locates right at the old peak of the model predicted distribution, closer to the result of Blanchard17 than that of Troja17.
%Though the age of NGC 4993 given by Blanchard17 and Troja17 differs a lot from each other, we give our prediction on the most likely age of NGC 4993 from galaxy formation view. 
$95\%$ of the selected NGC 4993-like galaxies have little star formation. Their median sSFR is $10^{-13}\peryr$, as shown by the orange line in the lower panel of Fig. \ref{fig:other}. 

\section{Conclusion}
\label{section:conclusion}
In this work, we use the semi-analytic model of galaxy formation {\CODE} which includes modeling of binary evolution by adopting {\YunnanII} stellar population synthesis model to derive the neutron star-neutron star merger (NNM) and neutron star-black hole merger (NBM) event rates for different kinds of galaxies. 
%As natal kick velocity of supernovae $\Vkick$ has huge influence on NNM events rate for simple stellar populations, we set $\Vkick$ to be a free parameter, from $0$ to $190\kms$. We also mark NNMs which have merger timescale longer than $6\Gyr$ as ``Old'' NNMs, as the analogue of GW170817, which most likely have merger timescale longer than $6.8\Gyr$. 
After presenting the NNM and NBM event rates in different simple stellar populations (SSPs) predicted by the {\YunnanII} model, we study the predicted cosmic NNM and NBM event rate density, r-process elements produced through these mergers, and the properties of host galaxies of the mergers. Here are the main results:
\begin{itemize}
    \item In {\YunnanII} stellar population synthesis model that models binary evolution, the value of natal kick velocity of supernovae $\Vkick$ assumed in the model affects the NNM rates in SSPs, and also affect the cosmic NNM rate density, by as much as one magnitude, when changing the value from $190\kms$ to $0$. The cosmic NNM rate density predicted with $\Vkick=0\kms$ (which we choose as the fiducial model) fits the observational result of LIGO Scientific Collaboration and Virgo Collaboration (LVC) best. However, the observation of single pulsar proper motions claims a much larger kick velocity ($190\kms$, \citealt{Hansen97}). This inconsistency may indicate that the evolutionary tracks and kick velocities of neutron stars in binary systems could be different from single neutron stars, and a more refined and physical model for kick velocities in binary evolution may be needed. Note that we have not done the full exploration of parameter space, which is beyound the scope of this work. Thus this result should be treated with caution. In our model, NNMs prefer to originate from binary systems with low kick velocities.
    %The inconsistency between the observational constraint of single pulsar kick velocities ($190\kms$, \citealt{Hansen97}) and our model preference for low kick velocities ($<50\kms$) in the progenitors of NNMs indicates that a more refined and physical model for kick velocities in binary evolution is needed, and may also be alleviated by more GW detection in the future.
    %which supports the theoretical expectation that natal kick velocities in binary evolution should be smaller than that in single star case. 
    Whereas, the NBM event rate density is almost independent of value of kick velocity in our model.\\
    \item The predicted cosmic NNM events rate density at $z=0$ of our fiducial model is $283\Gpcyr$, marginally in agreement with the value constrained by LVC GWTC-1 ($1210_{-1040}^{+3230}\Gpcyr$). The NNMs that have similar old age as GW170817 are about $30\%$ of all NNMs at $z=0$. We expect that more NNMs in young galaxies should be observed in the future. The predicted cosmic event rate density of NBMs at $z=0$ of our fiducial model is $91\Gpcyr$, about a third of the one of NNMs, which is also consistent with the upper limit proposed by LVC GWTC-1 ($610\Gpcyr$). \\%The ratio of observed NNMs and NBMs should be around $3:1$.\\ %in the future, otherwise modifications of stellar evolution models would be needed.\\
    \item The predicted total number of NNMs and NBMs in the whole life of a Milky Way-mass galaxy is $2.0^{+0.5}_{-0.4}\times10^6$. By assuming yield mass $\MejectaNNM=\MejectaNBM=0.01\Msun$, the corresponding amount of r-process elements with $A>79$ is $2.0^{+0.5}_{-0.4}\times10^4\Msun$, comparable to the observational constraint ($1.90^{+0.22}_{-0.21}\times10^4\Msun$). %Nevertheless, the yield mass of $0.04\Msun$ overestimates the amount.  The constrained range of ejecta from GW170817 is $0.01-0.04\Msun$, which means NNMs alone or NNMs plus NBMs can provide the amount of r-process elements we observe, and may provide more than we need.
    Milky Way-mass galaxies are the main sites for historical r-process nucleosynthesis.\\
    \item For a Milky Way-mass galaxy at z=0, the predicted NNM rate is $25.7^{+59.6}_{-7.1}\Myr^{-1}$, in a good agreement with the observational result of the Milky Way ($21^{+28}_{-14}\Myr^{-1}$, \citealt{Kim15}). %The predicted NBM rate in the Milky Way-mass galaxies is $2.1^{+40.3}_{-1.8}\Myr^{-1}$, with larger scatter than the one of NNM.\\
    In general, young and massive galaxies have higher NNM and NBM rate. %The age dependence mainly comes from young NNMs, not Old ones. The rate of Old NNMs almost has no dependence on age. 
    %The stellar mass and metallicity distribution of host galaxies of NNMs and NBMs are unimodal. 
    NNMs and NBMs are most possible to be detected in galaxies with $\Mstar\sim10^{10.65}\Msun$ and metallicity of $12+\log{\rm (O/H)} = 8.72-8.85$, %The age, color, specific star formation rate (sSFR) and morphology distribution of host galaxies are bimodal. 
    and are more in young, blue, star-forming and disk galaxies. %The detailed probabilities of detection in different kinds of galaxies is listed in Tab. \ref{tab:other}. 
    The properties of NGC 4993, the host galaxy of GW170817, are mostly at or near the peaks of model predicted distrbutions, %(except age, but this inconsistency is caused by the huge uncertainties in the determination of age through spectrum)
    indicating that NGC 4993 is a typical host galaxy for NNMs.%, and it is not surprising to detect the first NNM in such a galaxy.
\end{itemize}

During LVC O1 and O2, only one NNM event and corresponding host galaxy was detected. LVC O3 began on 1st April, 2019 and is planned to end on 30th April, 2020. Three NNMs and two NBMs (probability $>99\%$) have been detected till the end of 2019. Though not all the electromagnetic counterparts can be confirmed, the growth of event number is very inspiring. With larger observational sample coming in the future, we can switch the study mode from case study to statistics. The observational distributions of binary compact objects and host galaxies' properties can be used to constrain all the physical models involved, helping us have a better understanding of stellar evolution, compact objects and galaxy formation.

\section*{Acknowledgements}
We acknowledge Hailiang Chen for reading our draft and providing some very useful comments and suggestions. J.W. acknowledges the support from the National Natural Science Foundation of China (NSFC) grant 11873051. F.Z. is supported by NSFC grants 11573062, 11973081, 11521303, the YIPACAS Foundation grant 2012048, and the Yunnan Foundation grant 2011CI053. L.X.L. is supported by NSFC grants 11373012, 11973014. R.L. is supported by NSFC grants 11773032, 118513, and the NAOC Nebula Talents Program. L.G. is supported by the National Key R\&D Program of China (NO. 2017YFB0203300), and the Key Program of NFSC through grant 11733010. Z.H. is partly supported by NSFC grants 11521303, 11733008. J.P. acknowledges support from the National Basic Research Program of China (program 973 under grant no. 2015CB857001).

%We acknowledge support from the National Key Program for Science and Technology Research and Development (2015CB857005, 2017YFB0203300) and the Chinese Natural Science Foundation (NSFC) grants (11390372...to be finished, Z.H. is partly supported by the Natural Science Foundation of China (Grant Nos 11521303,11733008).

%%%%%%%%%%%%%%%%%%%%%%%%%%%%%%%%%%%%%%%%%%%%%%%%%%



\bibliographystyle{mnras}
\bibliography{NSNS} % if your bibtex file is called example.bib

%\begin{thebibliography}{99}
%\end{thebibliography}

%%%%%%%%%%%%%%%%%%%%%%%%%%%%%%%%%%%%%%%%%%%%%%%%%%

%%%%%%%%%%%%%%%%% APPENDICES %%%%%%%%%%%%%%%%%%%%%

\appendix

\section{Galaxy number density distribution in GABE}
\label{section:numberdensity}

\begin{figure}
  \centering
  \includegraphics[width=0.5\textwidth]{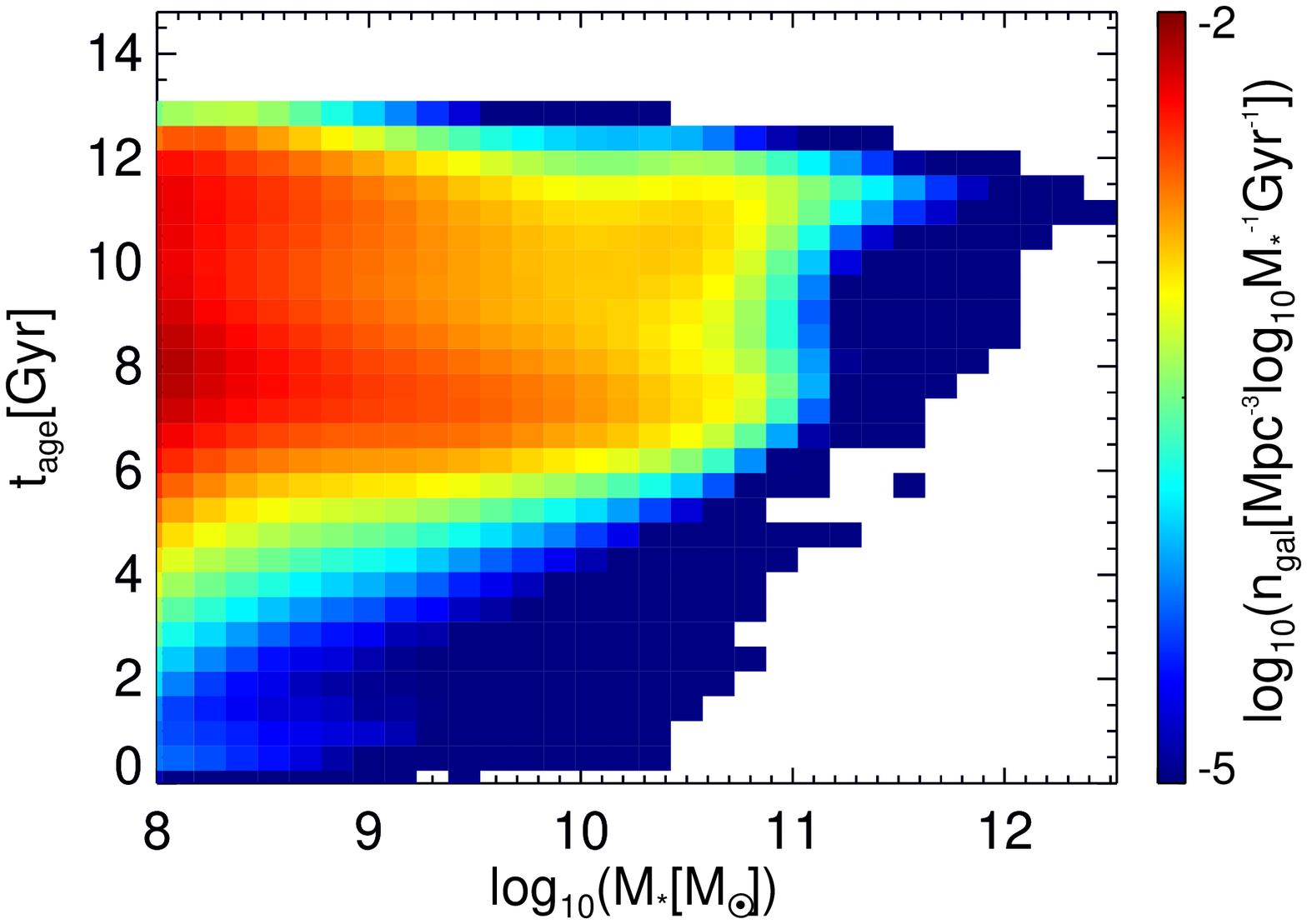}
  \caption{The number density distribution of galaxies as a function of stellar mass and mass-weighted age at $z=0$ in {\CODE}. Color in each cell indicates the mean number density for galaxies in the cell with values shown in the color bar.}
  \label{fig:agedis}
\end{figure}

Fig. \ref{fig:agedis} presents the spacial number density distribution of all galaxies at $z=0$ in {\CODE}, in a age -- stellar mass plane. The distribution peaks at stellar mass $\sim 10^8 \Msun$ and age $\sim 8 \Gyr$. Benefit from the inclusion of almost all classical galactic physical processes, {\CODE} provides us a fair complete star formation history library for the calculation of $\RCOM$. Other properties of galaxies in {\CODE} can be found in \cite{Jiang19}.
%%%%%%%%%%%%%%%%%%%%%%%%%%%%%%%%%%%%%%%%%%%%%%%%%%

% Don't change these lines
\bsp	% typesetting comment
\label{lastpage}
\end{document}